\documentclass{aa}  

\usepackage{graphicx}
\usepackage{txfonts}
\usepackage{comment}
\usepackage{booktabs}
\usepackage{float}
\usepackage{enumitem}
\usepackage{xurl}
\usepackage{hyperref}

\begin{document} 

   \title{Thermophysical characterization of the potentially hazardous near-Earth object 2024 YR4}

   \author{T. G. M\"uller\inst{1}
      \and E. M. MacLennan\inst{2}
      \and A. Y. Burdanov\inst{3}
      \and B. J. Holler\inst{4}
      \and A. S. Rivkin\inst{5}
      \and J. de Wit\inst{3}
      \and P. Pravec\inst{6}
      \and M. Micheli\inst{7}
      \and K. Muinonen\inst{2}
      \and D. Farnocchia\inst{8}}

    \institute{Max-Planck-Institut für extraterrestrische Physik (MPE), Giessenbachstraße 1,
              D-85748 Garching, Germany\\ \email{tmueller@mpe.mpg.de}
          \and
              Department of Physics, University of Helsinki, P.O. Box 64, FI-00014, Helsinki, Finland
          \and
              Department of Earth, Atmospheric and Planetary Sciences, Massachusetts Institute of Technology, 77 Massachusetts Avenue, Cambridge, MA 02139, USA
          \and
              Space Telescope Science Institute, Steven Muller Building, 3700 San Martin Drive, 21218 Baltimore, MD, USA
          \and
              Johns Hopkins University Applied Physics Laboratory, 11100 Johns Hopkins Road, Laurel, MD, 20723, USA
          \and
              Astronomical Institute, Academy of Sciences of the Czech Republic, Ond\v{r}ejov, CZ-25165, Czech Republic
          \and
              ESA NEO Coordination Centre, Largo Galileo Galilei, 1, 00044 Frascati (RM), Italy
          \and
              Jet Propulsion Laboratory, California Institute of Technology, 4800 Oak Grove Dr., Pasadena, CA 91109, USA
              }

   \date{Received; accepted}

\abstract
{The asteroid 2024 YR$_4$ is a potentially hazardous Apollo-type near-Earth object. Its highly eccentric orbit produces close Earth-Moon
encounters roughly every 4 years. Based on the astrometry collected during the discovery apparition, the asteroid had a 4\% probability
of a lunar impact on 2032 December 22.
A collection of light curves, including JWST-NIRCam data, allowed us to constrain the rotation period, spin pole, and 
convex shape solutions for the object.
We analyzed two sets of JWST-MIRI observations obtained in three filters on 26 March 2025.
The extracted photometry reveals rotationally driven thermal flux variations.
The combination of a spin-shape model from light-curve inversion and the absolutely calibrated three-band mid-infrared light curves
is well reproduced by a convex spin-shape solution with a spin pole at
($\lambda_{ecl}$, $\beta_{ecl}$) = (232 $\pm$1$^{\circ}$, -11 $\pm$3$^{\circ}$), a sidereal period of 19.4633 ($\pm$ 0.0001)\,min,
and axis ratios of $a$/$b$ $\approx$ 1.28 and $b$/$c$ $\approx$ 1.45.

The corresponding equivalent-volume diameter is 60.8$^{+1.8}_{-3.6}$\,m, with a geometric V-band 
albedo of $p_V = 0.11^{+0.05}_{-0.03}$ (assuming H$_{V}$ = 24.14 $\pm$0.25\,mag).
A high thermal inertia $\Gamma$ $>$500\,J\,m$^{-2}$\,K$^{-1}$\,s$^{-1/2}$ and low surface roughness are required
to explain the observed MIRI data. The rapid rotation, in combination with the high thermal inertia,
leads to significant heat transport to the night side, producing a broad nearly isothermal equatorial
band.

The near-Earth asteroid thermal model has limited reliability for this object, requiring
an extreme beaming parameter $\eta > 3$,
whereas the fast-rotating model and a full thermophysical model provide consistent and reliable estimates of the size and albedo.

The combination of high eccentricity and high thermal inertia implies seasonal skin depths of up to $\sim$3\,m and radiative
timescales of weeks to months. Although seasonal heating is negligible for the March 2025 observations because of the
near-equator-on viewing geometry, the seasonal Yarkovsky drift is expected to be dominant over the diurnal component and
might become measurable during future encounters.

Although JWST astrometry from February 2026 rules out any possible impact on the Earth or the Moon for at least 100 years,
the spin, shape, and radiometric constraints
presented here provide an important benchmark for future potentially hazardous objects.}

\authorrunning{M\"uller et al.}
\titlerunning{TPM Characterization of NEA 2024~YR4}

\maketitle

\section{Introduction}
\label{sec:intro}

The near-Earth asteroid 2024~YR$_4$ (YR4 in short) was discovered by the ATLAS\footnote{\href{https://atlas.fallingstar.com/}{Asteroid Terrestrial-impact Last Alert System}} telescope
in Chile on 2024 December 27\footnote{\href{https://minorplanetcenter.net/mpec/K24/K24YE0.html}{MPEC 2024-Y140 : 2024 YR4}},
two days after a close encounter with Earth (on 25 December 2024 04:45 UTC at
2.16 lunar distances).
YR4 has a highly eccentric orbit ($a = 2.52$\,au, $i = 3.41^{\circ}$, $e = 0.66$) with perihelion and aphelion distances
of 0.85\,au and 4.18\,au, respectively. Regular encounters with the Earth-Moon system (e.g., December 2024, 2028, and 2032)
occur due to its current orbital period of 3.99\,yr.
With the high eccentricity and a semimajor axis close to the Jupiter 3:1 mean-motion resonance (MMR),
the orbit is dynamically unstable, with a short typical lifetime of $\sim$10$^{6}$\,years \citep{Gladman1997}.
The 3:1 MMR, the frequent close approaches to the Earth-Moon system, and non-gravitational forces
such as the Yarkovsky effect contribute to the growing uncertainty in the future trajectory of YR4,
making very long-term predictions difficult \citep{Farnocchia2026}.

Orbit calculations during the first weeks after discovery indicated a nonzero impact probability
with Earth on 2032 December 22. Additional astrometric measurements in February 2025 significantly improved the orbit solution,
ruling out an Earth impact in 2032,
but resulted in a 4.3\% chance of impact with the Moon\footnote{\href{https://science.nasa.gov/blogs/planetary-defense/2025/06/05/nasas-webb-observations-update-asteroid-2024-yr4s-lunar-impact-odds/}{NASA's Webb Observations Update Asteroid 2024 YR4’s Lunar Impact Odds}, retrieved
on 18 February 2026} \citep[see also][]{Micheli2026}. Such a lunar impact would be expected to create a $\sim$1~km crater, and an estimated $\sim$10$^{8}$\,kg of ejecta material would
escape lunar gravity \citep{He2026}. This material could cause severe problems for lunar exploration activities, for satellites in near-Earth space, and depending on 
the impact location on the Moon, up to 1000\,tons of material might accrete on the Earth on timescales of a few days \citep{Wiegert2025}, which would also
create a lunar meteor outburst within 100 years \citep{He2026}. However, in recent astrometric studies \citep{Rivkin2026,Deen2026,deWit2026},
an impact on the Moon in 2032 was finally excluded.

The size estimates from the object's V-band absolute magnitude, $H_{V}$, of 24.14 $\pm$ 0.25\,mag
\citep{Devogele2026} range between 25\,m (very high albedo p$_V$ = 0.50) to 128\,m
(very low geometric albedo of p$_V$ = 0.03).
\citet{Bolin2025} obtained multiband visible and near-infrared (NIR) measurements and determined a
spectral slope $g - i$ of 13\% $\pm$ 3\%/100\,nm and color indices $g - r$, $r - i$, $i - Z$, and $Y - J$, indicating a similarity to R- or Sa-type asteroids. Based on this analogy, \citet{Bolin2025} estimated a size of $\sim$30-65\,m (assuming a most likely intermediate to high albedo
of 0.15-0.4). Their light-curve data, showing a variation of $\sim$0.4 mag, indicated an oblate shape with a $\sim$3:1 axial ratio with a retrograde spin with
a double-peaked synodic rotation period of 1172 $\pm$ 284\,s and a spin pole at ecliptic longitude, latitude coordinates
($\lambda$, $\beta$) = ($\sim$42$^{\circ}$, $\sim$-25$^{\circ}$).

To estimate the related risk in case of an impact, a more accurate physical description of YR4 was needed.
Thermal (or mid-)infrared (MIR) measurements were
chosen to obtain a radiometric size estimate \citep{Delbo2015,Mueller2014}. However, in late January 2025, when the object had reached
the Torino level 3 classification\footnote{\href{https://www.planetary.org/articles/the-torino-scale}{The Torino Scale}},
YR4 was already very faint in the MIR, well below 1\,mJy in N band, and beyond reach of MIR facilities on the ground (IRTF\footnote{\url{https://irtfweb.ifa.hawaii.edu/}} or
VLT\footnote{\url{https://www.eso.org/}}), 
with the James Webb Space Telescope (JWST\footnote{\url{https://science.nasa.gov/mission/webb/}}) being the only
remaining possibility for a detection in the MIR in 2025.

We present Gemini North observations and the JWST NIRCam and MIRI observations (Sect.~\ref{sec:obs}), including data processing, calibration, and photometric extraction
steps. Based on the available light-curve measurements, we determined a range of possible spin-shape solutions (Sect.~\ref{sec:spinshape}).
In Sect.~\ref{sec:radiometric} we apply different model concepts to constrain the object's radiometric and size properties.
The discussion of the results and the conclusions are presented in Sect.~\ref{sec:dis} and Sect.~\ref{sec:con}, respectively.

\section{Observations}
\label{sec:obs}

\subsection{Gemini North Observations}

Quasi-simultaneous visible-wavelength ground-based monitoring of YR4 was performed to aid in constraining the overall spin-shape state of the asteroid. Observations with the 8.1\,m Gemini North telescope (MPC code: T15) were conducted on 6 March 2025 from 09:25 to 10:14 UTC as part of the program GN-2025A-DD-103 (PI: Burdanov) when the asteroid was located at a heliocentric distance $r$ = 1.627\,au, at a distance $\Delta$ = 0.760\,au from the Earth, and at a phase angle of $\alpha$ = -25.0$^{\circ}$ (trailing the Sun). Images were obtained with the Gemini Multi-Object Spectrograph (GMOS; \citealt{2004PASP..116..425H}) in imaging mode in the broadest available filter to maximize the flux from YR4 (\textit{ri} filter covering 560-850\,nm). 

Observations were performed using non-sidereal tracking with 2$\times$2 binning, corresponding to 0.16\,arcsec\,pixel$^{-1}$, and a 120\,s exposure time. The exposure time was chosen to increase the signal-to-noise ratio (SNR) while avoiding commensurability with the object's rotation period (i.e., to avoid sampling the same rotational phase) and to properly sample the fast-rotation period. The asteroid was placed on the central chip of the GMOS sensor arrays. Observations spanned an air mass from 1.1 to 1.3, sub-arcsec seeing ($\sim$0.8\,arcsec), and resulted in 21 images. 

The images were calibrated using the pipeline DRAGONS \citep{2023RNAAS...7..214L}, which applies standard Gemini GMOS data reduction procedures. The calibration includes subtraction of a master bias frame and over-scan correction, followed by trimming of the detector edges and flat-field correction using a master flat to account for pixel-to-pixel sensitivity variations. The data were then converted from ADUs to electrons, and corrected for relative quantum efficiency differences between the CCDs. The individual detector chips were subsequently mosaicked into a single image. The reduced images were then astrometrically solved using Astrometry.net \citep{2010AJ....139.1782L} and aligned such that the asteroid moved across the detector while background stars remained stationary, exhibiting slight trailing due to the initial non-sidereal tracking during the observations. Aperture photometry was performed in Python using a growth-curve approach\footnote{\href{https://photutils.readthedocs.io/en/latest/index.html}{Astropy Photutils}} \citep{Bradley2025} to determine the optimal aperture size (approximately 4 pixels in radius) that maximized the SNR ($\approx$5 on individual images), with field stars from the Gaia catalog used as photometric references. The results are listed in Table~\ref{tbl:photometry_mag}.

\subsection{JWST observations}

A director's discretionary time (DDT) observing program on the JWST was set up (DD 9239; \citealt{Rivkin2025jwst}),
including astrometric measurements for orbit improvements with
the Near-Infrared Camera (NIRCam; \citealt{Rieke2023}) and thermal measurements with the Mid-Infrared Instrument (MIRI; \citealt{Rieke2015a})
enabling the only size measurements that would be possible until 2028.
JWST NIRCam measurements (in F150W2 and F322W2 bands) were executed on 8 and 26 March and on 11 May 2025. The planned MIRI observations
on 8 March and 11 May 2025 were skipped due to guide-star problems, and only two MIRI measurements (each time in the F1000W, F1280W, and F1500W bands)
were conducted on 26 March 2025.

\subsubsection{JWST-NIRCam measurements in March 2025}
\label{sec:miri_obs}

\paragraph{Observations}

The JWST Science Archive\footnote{\href{https://jwst.esac.esa.int/archive/}{ESA JWST Science Archive}}
contains a list of NIRCam \citep{Rieke2023} data of YR4 taken in 2025 (Proposal ID: 9239) and 2026 (proposal ID: 9441).
Observations in two filters, F150W2 (short, $\Delta \lambda$ = 1.228\,$\mu$m) and
F322W2 (long, $\Delta \lambda$ = 1.340\,$\mu$m), were taken simultaneously.

Photometry in the two filters from 8 March was obtained, but only the F150W2 measurements for 26 March were usable
due to the diminished brightness of YR4 in the long-wavelength filter. The measurements on 11 May 2025 (and
on 18 and 26 February 2026 were only used for astrometric studies and the SNR was too low for 
extracting light-curve information (the estimated visual magnitudes were about
25.6, 26.6, 28.3, and 30.1\,mag on 8 and 25 March 2025, on 11 May 2025, and in February 2026, respectively).
The photometric results are listed in Table~\ref{tbl:short_long_fluxes}.

\subsubsection{JWST-MIRI imaging on 26 March 2025}

\paragraph{Observations}

We obtained MIR measurements with
MIRI \citep{Rieke2015a,Wright2015} in imaging mode \citep{Bouchet2015},
as part of the JWST cycle 3 proposal "Size Measurements of a Potential Earth-Impacting Asteroid with
JWST MIRI and NIRCAM" (ID.\ \#9239; see also \citealt{Rivkin2025jwst}).
The data comprise two multi-band sequences with the F1280W ($\Delta \lambda$ = 2.47\,$\mu$m),
F1000W (1.80\,$\mu$m), and F1500W (2.92\,$\mu$m) filters,
taken on 26 March 2025, with a total of about two hours integration time, but spread
over roughly four hours.
Details of the measurements are
listed in Table~\ref{tbl:miri_obs} and in Table~\ref{tbl:miri_obs56_indint} .
We give the precise start and end times (in the JWST reference frame) of each 
measurement. For a switch to the object's reference frame,
the light-travel time would have to be subtracted (between 8.949 and 8.974 minutes, which is slightly more than half
a rotation period).

\begin{table*}[h!tb]
  \caption{Results of the MIRI imaging-mode observations of asteroid 2024~YR$_{4}$, taken on 26 March 2025. \label{tbl:miri_obs}}
  \centering
    \begin{tabular}{lcrrrrrrr}
      \hline \hline
       Observation & Filter & EXPSTART\tablefootmark{a} & EXPMID\tablefootmark{a} & EXPEND\tablefootmark{a} & EFFEXPTM\tablefootmark{b} & SNR\tablefootmark{c}    & flx\tablefootmark{c}       & err\tablefootmark{c} \\
       dataset    & /Band  & (MJD)    & (MJD) & (MJD)  & (s)      &        & ($\mu$Jy) & ($\mu$Jy) \\
      \hline
      \noalign{\smallskip}
      OBS~5 \& OBS~6             & F1280W & 60760.21756 & 60760.28110 & 60760.34465 &  2464.2 & 36 & 6.395 & 0.198 \\
      combined\tablefootmark{d,e}  & F1000W & 60760.24116 & 60760.30474 & 60760.36832 &  2464.2 & 36 & 3.060 & 0.095 \\
                                 & F1500W & 60760.26493 & 60760.32852 & 60760.39212 &  2464.2 & 40 & 9.355 & 0.269 \\
      \noalign{\smallskip}
      OBS~5                      & F1280W & 60760.21756 & 60760.22779 & 60760.23805 &  1232.1 & 26 & 6.26  & 0.25  \\
      (05:13...06:51 UTC)        & F1000W & 60760.24116 & 60760.25140 & 60760.26166 &  1232.1 & 25 & 2.77  & 0.12  \\
                                 & F1500W & 60760.26493 & 60760.27521 & 60760.28549 &  1232.1 & 28 & 9.26  & 0.35  \\
      \noalign{\smallskip}
      OBS~6                      & F1280W & 60760.32416 & 60760.33441 & 60760.34465 &  1232.1 & 29 & 6.29  & 0.23  \\
      (07:46...09:24 UTC)        & F1000W & 60760.34783 & 60760.35807 & 60760.36832 &  1232.1 & 30 & 3.17  & 0.11  \\
                                 & F1500W & 60760.37154 & 60760.38182 & 60760.39212 &  1232.1 & 28 & 8.95  & 0.34  \\
      \hline
    \end{tabular}
      \tablefoot{The observations were taken as part of the cycle~3 DDT proposal ID. 9239,
                 OBS~1 to OBS~4 are related to NIRCam measurements, which are not listed here.\\
      \tablefoottext{a}{The modified Julian day (MJD) times (EXPSTART/-MID/-END) are taken from the
          data product FITS headers, and they correspond to the JWST spacecraft reference time frame.}
      \tablefoottext{b}{The effective exposure time (EFFEXPTM) is the total time for signal
          accumulation and relevant for the noise calculations.}
      \tablefoottext{c}{The signal-to-noise ratio
          (SNR), the fluxes (flx) and errors (err) are obtained via aperture photometry; the errors include the formal absolute flux
          calibration error for MIRI imaging.}
      \tablefoottext{d}{The combined data (per band) consist of an integration time of two times 29.5\,min (corresponding to about
                        $2x 150$\% of the object's rotation period), but with a gap of about 124\,min in between).}
      \tablefoottext{e}{At the midpoint of the observation sequence at 7:18 UTC, the object was 1.8106\,au from the Sun and
      1.0775\,au from JWST, and it was seen under a phase angle of -28.5$^{\circ}$. For our radiometric study, we took a signed
      version of the phase angle $\alpha$. It is needed to correctly handle the heat transport from the illuminated to
      the non-illuminated part of the surface for a given spin vector of the object.}
                }
\end{table*}

The first MIRI multi-filter observing block (OBS~5) started on 26 March 2025 at 05:13 and ended at
06:51 UTC. The second one (OBS~6) started at 07:46 and ended at 09:24 UTC, followed by a sequence
of NIRCam measurements on the same target (OBS~2, OBS~7, OBS~8).
Each block contains measurements using the filter sequence F1280W, F1000W, and F1500W.
For all three filters, MIRI exposures were taken in each of four dither positions, with
each exposure composed of 4 integrations with 27 groups each (non-destructive reads covering
2.775\,s of integration).
The intention was coverage of more than a full rotation period before switching filters.

\paragraph{Data reduction and calibration}

The Level 1 {\tt uncal} files were retrieved from the Mikulski Archive
for Space Telescopes (MAST) hosted by STScI and processed locally using
version 1.18.0 of the JWST calibration pipeline \citep{Bushouse2025}.
We ultimately created a range of pipeline-processed images in each of the
three MIRI imaging filters, with the {\tt jump} step (i.e., the cosmic
ray rejection step) turned on and the {\tt clean\_flicker\_noise} step
turned off in each case:

\begin{itemize}
\item \textbf{OBS5 and OBS6 combined}, having a total integration time
  of about 3545\,s. These products were similar to the separate OBS5
  and OBS6 products, except that all dithers (each about 308\,s) from the two observations
  were combined, further increasing the SNR of
  the asteroid, and covering about three full rotations of the object.
\item \textbf{OBS5 and OBS6 separated}, with an integration of about
  1772\,s each; here, all four dithers per observation were combined
  and an outlier rejection step was applied that removes large pixel
  values from the average. This has the effect
  of removing the star streaks, since the stars are not in the same
  positions on the detector in each integration, thus leaving behind only the asteroid
  in the center of the image. YR4's rotational variability was averaged out
  as a result of combining all dithers within an observation.
\item Images for \textbf{each individual integration}, covering about
  75\,s each. These products required the most pre-processing prior
  to running the JWST pipeline, since each integration within each dither
  was first broken out into a separate file and then run through the
  pipeline individually. The purpose of this was to increase the
  temporal sampling for constructing a MIR rotational light curve
  of the asteroid.
\end{itemize}

The photometric calibration takes the time-dependent reduction in observed count
rates\footnote{\href{https://jwst-docs.stsci.edu/jwst-calibration-status/miri-calibration-status/miri-imaging-calibration-status}{MIRI Imaging Calibration Status}}
into account by using a model of the evolving response function. In addition, there is
a response correction needed to bring the different sub-array measurements up to
the signals obtained via the FULL frame. \citet{Gordon2025} found
that these corrections push the formal MIRI imaging flux calibration uncertainties
down to 1\% or better.

The MIRI flux calibration is based on the assumption of F$_{ref}$ ($\lambda$) = const.
\citep{Gordon2022}. For objects with spectral energy distributions deviating from the
reference spectrum, one has to color-correct the in-band fluxes to obtain monochromatic
flux densities at the MIRI reference wavelengths (here, at 10.0, 12.8, and 15.0\,$\mu$m).
However, we find that in the case of YR4 these corrections would be only around 1\% in all three bands
(for a very wide range of object properties), and we safely ignored the correction
\citep[see also][]{Mueller2023}.

\paragraph{Photometry and astrometry}

YR4 is clearly visible in the various images (or can be found by translating the object R.A.\ and Dec.\
coordinates into pixel coordinates via the WCS given in the corresponding fits headers). We
performed aperture photometry for OBS5 \& OBS6 combined, and also for the individual OBS images,
by centering the aperture on the source and following the recommended aperture photometry recipe
(aperture sizes that encircle 80\% of the energy; \citealt{Gordon2025}).

For the final calculation of the flux errors, we considered the 1$\sigma$ noise level
from the image analysis and added quadratically a 1\% error for the MIRI calibration
scheme \citep{Gordon2025} and another 1\% for possible color errors in the MIRI bands (to account for
differences between the baseline calibration spectrum with F$_{ref}$ ($\lambda$) = const.\ and
the true object spectrum). The corresponding formal absolute flux errors (see last column
in Table~\ref{tbl:miri_obs}) are around 3\% for the combined OBS5 \& OBS6 measurements, and around 4\%
for the individual measurements. The errors are in agreement with findings on calibration star measurements
\citep{Dicken2024}. However, for our radiometric analysis and the $\chi^2$ calculations, we added
an estimated 3\% in quadrature to account for imperfect background elimination as YR4 moves on a
structured MIR background.

For the individual integration maps, we first had to do a
background subtraction (using a second-order polynomial fit to the background) to bring the
background level close to zero. In a second step, we subtracted a median average
(in the detector reference frame) of the three other dither images. This procedure eliminated
the warm and noisy pixels in close proximity to the source without affecting the final photometry.
For obtaining higher SNRs, we used smaller aperture sizes (60\% encircled energy)
with the corresponding aperture correction factors.
For a better placement of the apertures, we determined the source center (in pixels)
from PSF-filtered images. We obtained final SNRs of about 10-20 for
individual integration measurements, the flux errors are between 5 and 14\%
(see Table~\ref{tbl:miri_obs56_indint}).
For our radiometric study we added to the formal flux errors another 6\% in quadrature
to account for residual effects from the elimination of warm and noisy pixels, and for
uncertainties in centroiding the apertures.

The MIRI imaging minimum detectable flux densities (SNR = 10, in 10,000\,s integration time)
for unresolved point sources (in FULL array configuration and on a low sky background) are
listed as 0.46, 0.83, and 1.18 $\mu$Jy in the F1000W, F1280W, and F1500W filters, respectively
\footnote{\href{https://jwst-docs.stsci.edu/jwst-mid-infrared-instrument/miri-performance/miri-sensitivity}{MIRI Sensitivity}}.
Translating these numbers to effective exposure times listed above, leads to 1-$\sigma$ noise levels of
0.09 (2464.2\,s), 0.13 (1232.1\,s), and 0.53 $\mu$Jy (74.9\,s) in F1000W,
0.17,             0.24,             and 0.96 $\mu$Jy in F1280W, and,
0.24,             0.34,             and 1.36 $\mu$Jy in F1500W, respectively.
These ETC-based\footnote{\href{https://jwst-docs.stsci.edu/jwst-exposure-time-calculator-overview}{JWST Exposure Time Calculator}}
noise levels are close to the noise level determined from the aperture photometry in the OBS~5 and OBS~6 image products
(see SNRs in Table~\ref{tbl:miri_obs}).
For the individual integration maps (Table~\ref{tbl:miri_obs56_indint})
we find noise levels below the ETC values. This is mainly related to the background determination,
which includes multiple integrations from the four dither positions (and therefore corresponds
to a much longer effective integration time).
For our YR4 measurements we also benefited from a low absolute sky background (below 16\,MJy/sr
at 10\,$\mu$m) due to the large solar elongation of 120.8$^{\circ}$. At a later time, during the
end of the March-to-May visibility window at solar elongations around 85$^{\circ}$, the MIR
background would have been a factor of two higher, requiring significantly longer exposure times
to obtain the same SNRs.

We also compared JWST-centric coordinates of YR4 (produced by JPL/Horizons
\footnote{\href{https://ssd.jpl.nasa.gov/horizons/app.html}{JPL/Horizons}} at the mid-time of
each dither image, and with orbital elements as of July 2025) with the
pointing information as given in the FITS header keywords {\tt MT\_RA} and {\tt MT\_DEC}.
We found a mean pointing offset of 0.295 $\pm$ 0.011\,arcsec, corresponding to about 2.7 MIRI pixels.
This offset is caused by the uncertainty of YR4's orbital elements at
the time of the JWST observation planning (early 2025) in combination with the 
absolute pointing error: the JWST pointing accuracy after target acquisition
is listed with 0.10\,arcsec\footnote{\href{https://jwst-docs.stsci.edu/jwst-observatory-characteristics-and-performance/jwst-pointing-performance/jwst-pointing-accuracy}{JWST Pointing Accuracy}}, but we found a slightly higher value of 0.14\,arcsec 
for typical observations of moving targets.

\section{Spin-shape solutions from light-curve inversion}
\label{sec:spinshape}

\subsection{Observations and methods}

We used the full collection of
light curves, including those in \citet{Bolin2025}, these in \citet{Devogele2026}, and the Gemini North and JWST-NIRCam light curves presented
in Tables~\ref{tbl:photometry_mag} and \ref{tbl:short_long_fluxes}, to determine spin-shape solutions for YR4. A summary is given in Table~\ref{tbl:obs_summary}. This photometric dataset spans a time period starting on 28 December 2024 10:52 UT until 26 March 2025 10:03 UT (87.966 days) and consists of observations over a solar phase angle range from 7.4$^{\circ}$ to 35.0$^{\circ}$. Only light curves acquired in R filter or that could be reliably transformed into the R filter were treated as absolute photometry (i.e., using known color indices) and all others were treated as relative photometry. The magnitudes were normalized to a heliocentric and observer-centric distance of 1\,au, and times were adjusted for the light travel time to represent the time the light left the object.
The absolute photometry was useful for determination of the photometric phase function in the light-curve inversion (LCI) modeling described below, for which we used the $H,G_{12}$ system \citep{Muinonen2010}. A Lommel-Seeliger surface scattering model was used \citep{Muinonen2022} with the $H,G_{12}$ phase function \citep[see Equ.~10 in][]{Muinonen2020} to calculate the integrated brightness of the object.

\begin{table}[h!tb]
          \caption{Summary of the light curves we used for the shape and spin modeling. \label{tbl:obs_summary}}
  \centering
    \begin{tabular}{lllll}
      \hline\hline
      \noalign{\smallskip}
       Observatory & UT Date &           & ph.               & filter \\
       (IAU code)  & YYMMDD & $N_{pts}$ & (\tablefootmark{a}) & information \\
      \noalign{\smallskip}
      \hline
      \noalign{\smallskip}
      \multicolumn{5}{l}{\citet{Devogele2026}:} \\
      Steward (I52) & 241228    & 32  & r & clear filter \\
      Danish (W74)  & 250103/04 & 72  & a & R filter \\
      LDT (G37)     & 250107    & 21  & r & SDSS r filter \\
      VLT (X11)     & 250121    & 45  & a & transf. to R \\
      NOT (Z23)     & 250131    & 15  & a & R filter \\
      \noalign{\smallskip}
      \multicolumn{5}{l}{\citet{Bolin2025}:} \\
      Gemini S. (I11) & 250207 & 16 & r & r,i transf. to r \\
      \noalign{\smallskip}
      \multicolumn{5}{l}{This Work:} \\
      Gemini N. (T15) & 250306 & 18 & r & ri filter \\
      NIRCam/JWST      & 250308 & 12 & r & F150W2 \\
      NIRCam/JWST      & 250308 & 12 & r & F322W2 \\
      NIRCam/JWST      & 250326 & 10 & r & F150W2 \\
      \noalign{\smallskip}
      \hline
      \noalign{\smallskip}
    \end{tabular}
      \tablefoottext{a}{Photometry a: absolute photometry; r: relative photometry}
\end{table}

The LCI procedure presented in \citet{Muinonen2020} and \citet{Muinonen2022} was used to constrain shape and spin solutions. The spin period was reported to be $\sim$1172 $\pm$ 284\,s \citep{Bolin2025} and 19.46341 $\pm$ 0.00008\,min \citep{Devogele2026} and we adopt 19.5\,min as a starting value in the LCI. First, we search for likely spin poles using simple triaxial ellipsoids and arrive at two likely regions near ecliptic longitudes of 30$^{\circ}$ and 210$^{\circ}$ and ecliptic latitudes near zero. We use these as initial values for the convex shape inversion using a least-squares solver. The convex shape is represented using spherical harmonic functions with a maximum degree of 4, meaning that there are 24 coefficients that are used to describe the shape. 

Next, a set of so-called virtual observations are generated from the light-curve observations by adding random noise computed drawn from Gaussian distributions based on their uncertainties \citep{Muinonen2020}. We calculate 1000 of these virtual observations for each pole (2000 in total) and then fit a shape to each set via a least-squares solver. The resulting distribution of virtual solutions can be used to estimate the uncertainty in the spin parameters. These virtual observations and solutions are then used in a Bayesian framework to serve as \emph{a priori} probability density functions \citep{Muinonen2020}, which represents the observational and model uncertainties. These probability functions were then fed into a Markov chain Monte Carlo (MCMC) setup, from which 500 spin shape solutions were generated for each of the two pole regions.

\subsection{Results and analysis}

The initial least-squares fit to the data yielded two possible poles at ecliptic longitude and
latitude ($\lambda$, $\beta$) = (42$^{\circ}$, +9.1$^{\circ}$) (pole~1\footnote{(R.A., Dec) = (36.52$^{\circ}$, +24.07$^{\circ}$)})
and (228$^{\circ}$, +1.5$^{\circ}$) (pole~2\footnote{(R.A., Dec) = (22.97$^{\circ}$, -15.75$^{\circ}$)}).
From the 1000 solutions for each of the two spin pole regions, we calculate uncertainty regions
of 37.1 $\pm$ 8.4$^{\circ}$, -25.4$\pm$ 17.2 $^{\circ}$ for pole~1 and 226.3$\pm$11.7$^{\circ}$, 7.9$\pm$25.8$^{\circ}$ for pole~2.

\begin{figure}[h!tb]
\centering
\includegraphics[width=0.99\hsize]{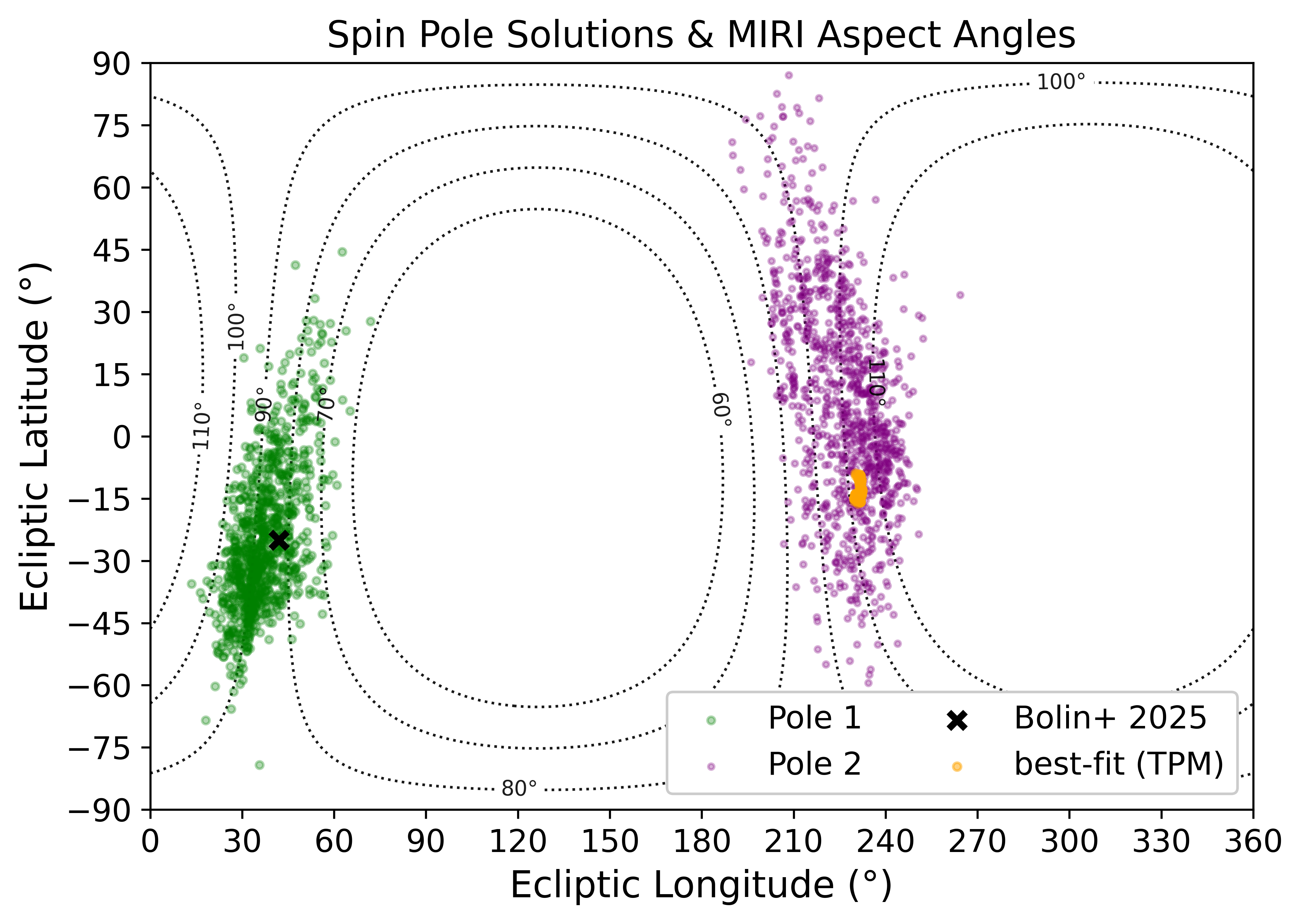}
\includegraphics[width=0.94\hsize]{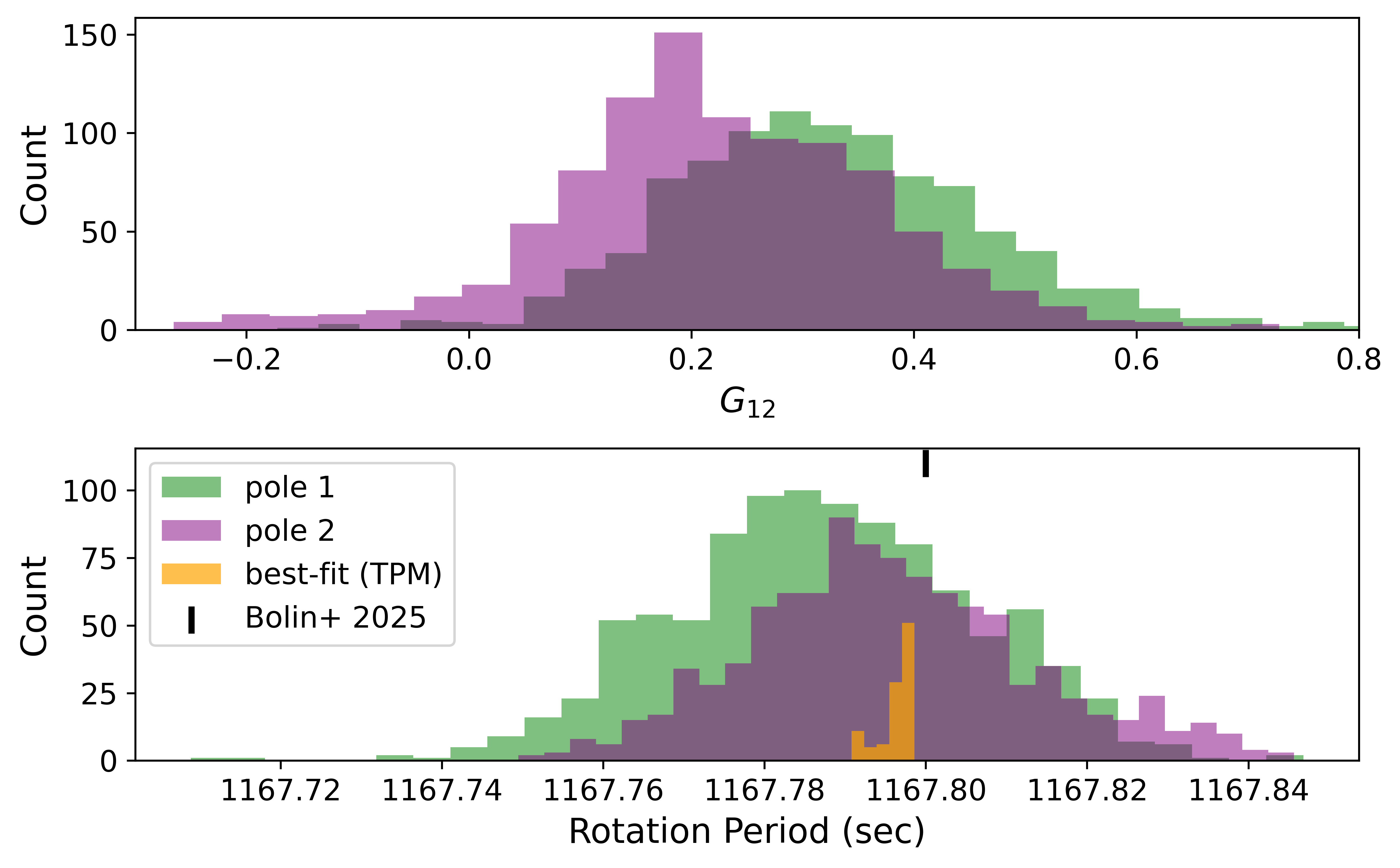}
\includegraphics[width=0.99\hsize]{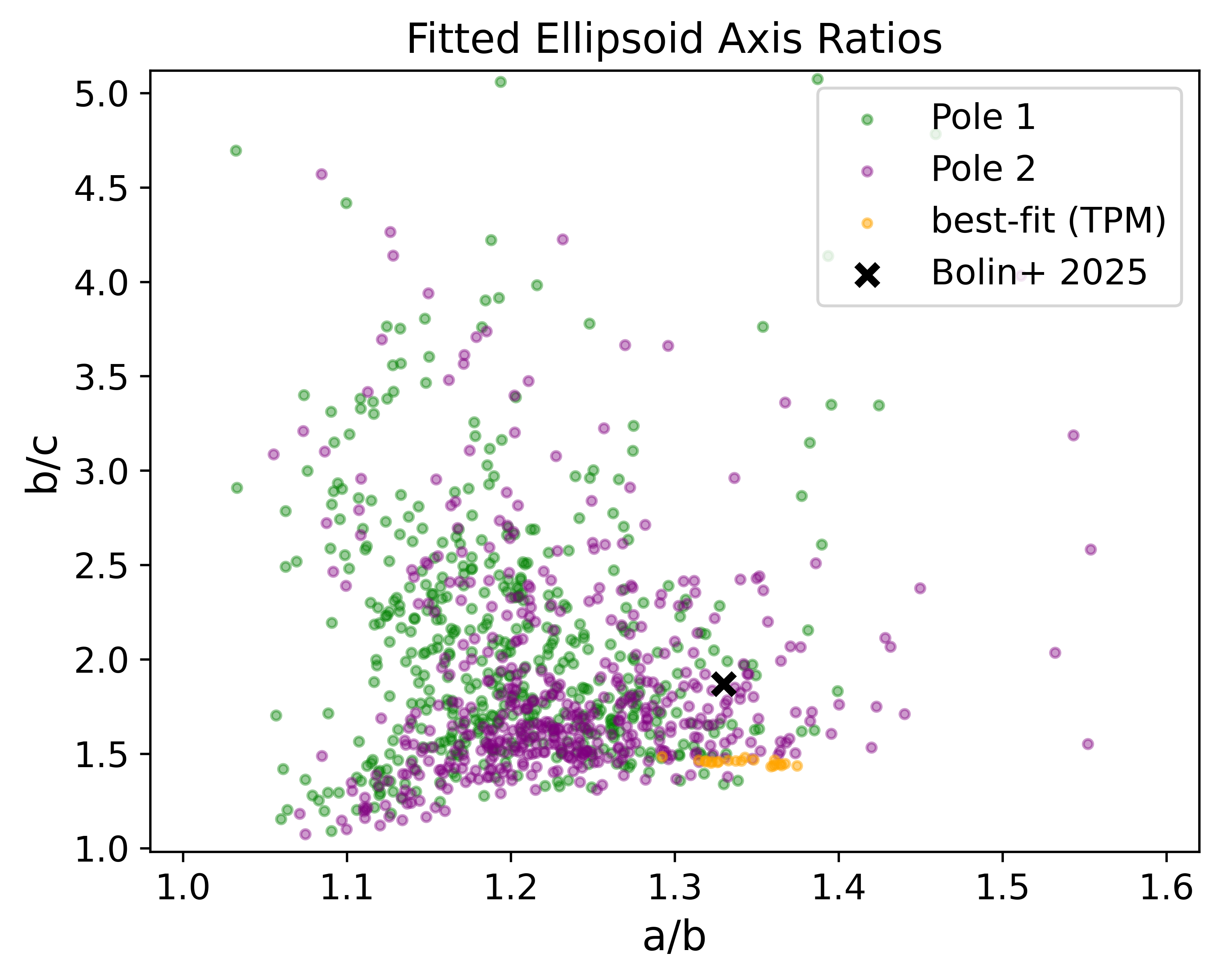}
   \caption{Constraints on the spin pole and axes ratios from the LCI
            analysis. Top: Spin-pole positions (in ecliptic longitude and latitude) of
            all light-curve-compatible solutions. The different colors indicate
            the range of pole~1 (green points) and pole~2 (purple points) solutions. The
            X stands for the \citet{Bolin2025} solution for comparison. Middle: Distributions of the $G_{12}$ and rotation period for each pole solution. Bottom: Constraints on the object's axial ratios $a$/$b$ and $b$/$c$ (as approximated by an equal-inertia ellipsoid) for the two possible islands of spin-pole solutions. The value for the \citet{Bolin2025} shape is again marked by X: $P_{sid}=1167.8$ s, $a$/$b$ = 1.33, $b$/$c$ = 1.87. The constraints from the MIRI three-band light curve (see Fig.~\ref{fig:allfluxes}) are shown in orange (see also the discussions in Sect.~\ref{sec:dis}).
      \label{fig:spinshape} }
\end{figure}

The solutions from the MCMC sampler initially returned shape solutions that, upon visual inspection, appeared too exotic. We therefore
re-started the MCMC 2-3 times to observe how the initial random selection of a spin-shape solution affected the result. We also tested some of these shapes in the TPM fitting
(see Sect.~\ref{sec:radiometric}) to MIRI data to assess reasonable shapes that also fit the thermal light curve. Our preferred spin-shape solutions for each pole are thus derived from this iterative process. In Fig.~\ref{fig:spinshape}, we show the posterior distribution of solutions from the preferred MCMC runs for each pole. We calculate the standard deviations of the spin pole, $G_{12}$ phase slopes, and rotation period distributions, as shown in the top and middle panels of Fig.~\ref{fig:spinshape}. Pole~1 is given by $46.0 \pm 2.7^\circ$, $-23.6 \pm 8.1^\circ$ and pole~2 is at $229.0 \pm 1.7^\circ$, $-9.7 \pm 5.1^\circ$. Note that these reported values are much smaller than those given for the virtual observations above.

The rotation period and estimated 1-$\sigma$ uncertainty for the pole~1 and pole~2 solutions are 1167.777 $\pm$ 0.002~s and 1167.794 $\pm$ 0.002~s, respectively. The 3-$\sigma$ uncertainty on the rotation period for each pole region is approximately 0.006\,s, meaning that the difference of 0.017~s is statistically robust to at least 7-$\sigma$ (Fig.~\ref{fig:spinshape}). These precise rotation periods owe to the large time span of the dataset, especially the NIRCam photometry, when compared to the rotation period of YR4.

As was done in \citet{MacLennan2026}, we fit a triaxial ellipsoid to each unique convex shape solution and compared their $a/b$ and $b/c$ axial ratios in the bottom panel of Fig.~\ref{fig:spinshape}. While these ratios are approximate representations of the convex shape, they are useful for selecting among the large set of shapes for fitting the MIRI thermal dataset.

The $H,\!G_{12}$ phase function was used for our shape model fitting. The two poles give consistent $G_{12}$ estimates: respectively, 0.23 $\pm$ 0.04 and 0.21 $\pm$ 0.04 for pole~1 and 2. For pole~2, we derive $G_1 = 0.22 \pm 0.04$ and $G_2 = 0.43 \pm 0.04$ \citep{Muinonen2010}. Although the $G_{12}$ parameter is not unique to any taxonomy, this range suggests a moderate albedo \citep{MacLennan2026,Pentikainen2026} and is thus consistent with the S-complex, K, and R taxonomic types that have been proposed for YR4 \citep{Bolin2025,Devogele2026}. 

Previous estimations of $H$ for YR4 have used different phase functions and with the assumption of a spherical shape, but with a large uncertainty of $\sim$0.2 mag. A significant source of uncertainty is in accurately approximating the opposition surge in brightness that happens at small phase angles. It was not possible to observe YR4 at these phase angles and so the uncertainty is dominated by the assumptions of these different phase function models \citep{Devogele2026}. It would be possible to revise the absolute magnitude using our set of shapes (in lieu of a spherical assumption), but the difference would be marginal and within the previously reported error bounds, which are dominated by systematic uncertainties.

\section{Radiometric analysis}
\label{sec:radiometric}

\subsection{Thermal model settings}
\label{sec:model_input}

We describe different models in Appendix~\ref{app:thermal_concepts}.
The standard thermal model \cite[STM,][]{Lebofsky1986, Lebofsky1989},
the near-Earth asteroid thermal model \cite[NEATM,][]{Harris1998},
the fast-rotating model \cite[FRM,][]{Lebofsky1978},
the isothermal model (ITM), and thermophysical model concepts \cite[TPM, see][and references therein]{Delbo2015}
allow us to determine the effective diameter of 2024~YR$_4$, as well as the surface temperatures and thermal properties.
Also thermal skin depth, radiation time scales, or the thermal parameter $\Theta$ are crucial for 
estimating the diurnal and seasonal effects.

Detailed radiometric studies require a range of input parameters, some of them are
known from other observations, others have to be assumed. We considered the
following parameters and ranges:

\begin{itemize}
\item The object's absolute magnitude in V band, $H_V$ = 24.14 $\pm$ 0.25\,mag, and the phase slope parameter $G$ = 0.51 $\pm$ 0.11 \citep{Devogele2026}.
\item The sidereal rotation period $P_{sid}$ = 19.46341 $\pm$ 0.00008\,min \citep{Devogele2026}, and slightly different values 
      connected to the multitude of spin-shape solutions from LCI techniques.
\item The JWST-centric observing geometry at mid-time (see Table~\ref{tbl:miri_obs}).
\item Available spin-shape models: spherical, ellipsoidal, spin-shape solution by \citet{Bolin2025}, own
      convex spin-shape solutions 
      based on the available light-curve measurements.
\item Assumed possible thermal inertia ranging from 0 to 2000\,J\,m$^{-2}$\,K$^{-1}$\,s$^{-1/2}$.
\item Assumed surface roughness: r.m.s.\ of surface slopes from 0$^{\circ}$ to 50$^{\circ}$ (from very low to very high roughness levels).
\item Emissivity $\epsilon_{bolo}$ = 0.9, and $\epsilon_{spec}$ = 0.9 (wavelength-independent).
\end{itemize}

\subsection{NEATM, FRM, ITM, and TPM: Spherical shape solutions}
\label{sec:spherical_results}

In a first analysis, we used the 10.0, 12.8, and 15.0\,$\mu$m fluxes listed under OBS~5 and OBS~6
in Table~\ref{tbl:miri_obs}. Each of the six individual measurements covers about 29.5\,min, which
corresponds to roughly 1.5 rotation periods of YR4. Rotational changes in cross-section are 
therefore averaged out and using spherical shape models is perfectly justified in the context
of radiometric studies.

\paragraph{NEATM, FRM, and ITM.}

\begin{figure}[h!tb]
\centering
\includegraphics[width=\hsize]{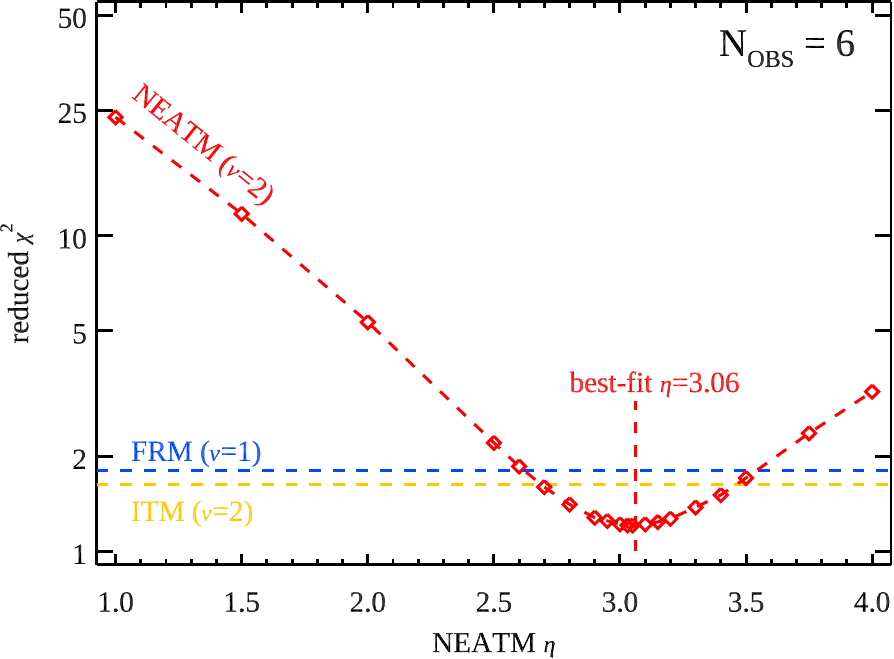}
\includegraphics[width=\hsize]{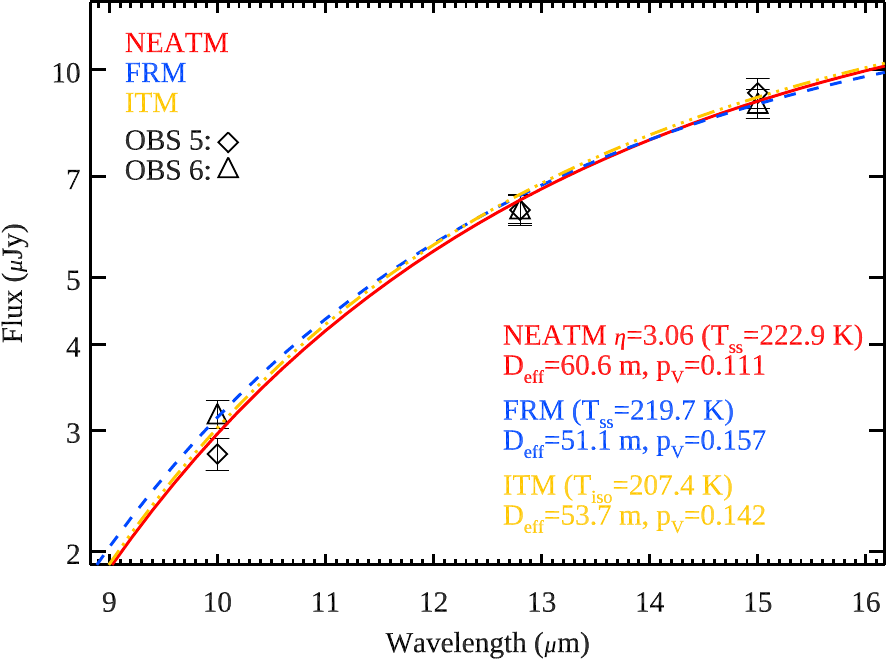}
   \caption{Fitting the observed OBS~5 and OBS~6 fluxes (after adding 3\% absolute flux error; see text)
            by the NEATM, FRM, and ITM for the JWST-centric observing geometry on 26 March 2025.
            Top: Reduced $\chi^2$-values for the three simple models. $\nu$ is the
            number of free parameters for the reduced $\chi^2$ calculations, and N is the number of measurements.
	    Bottom: OBS~5 and OBS~6 calibrated fluxes (shown as diamonds and triangles)
            together with the best-fit NEATM, FRM, and ITM solutions.
      \label{fig:ratios3}}
\end{figure}

We used the NEATM, FRM, and ITM setups (see description in the Appendix~\ref{app:thermal_concepts})
to obtain the best-fit solutions ($\chi^2$-metric) to the 6 individual fluxes and their absolute errors
(see also the description about the assumed flux uncertainties in Sect.~\ref{sec:miri_obs}).
Under the given JWST observing geometry (see Table~\ref{tbl:miri_obs}), the three models produce
different temperature patterns on the surface, in case of the NEATM, the temperatures also
depend on the assumed beaming parameter $\eta$. The matching of the model predictions
to the object's flux densities leads then directly to a size determination for the assumed
spherical shape. For the degrees of freedom $\nu$ in the $\chi^2$ calculations we used $\nu = 1$ for the
FRM and $\nu = 2$ for the ITM and NEATM (see Fig.~\ref{fig:ratios3}, top panel).

The NEATM requires a best-fit beaming parameter of 3.06 to obtain an acceptable reduced $\chi^2$
level well below 2.0. But also the FRM (reduced $\chi^2$ = 1.80) and also the ITM
(reduced $\chi^2$ = 1.63) can explain the MIRI measurements. However, the corresponding
object properties differ, depending on which model we used: (i) in case of NEATM, we find
a sub-solar temperature of around 223\,K, a size of 60.6\,m (diameter of the sphere), and
a geometric albedo of 0.11 (for the above listed H-mag); (ii) the sub-solar temperature in
the FRM is more than 3\,K lower (219.7\,K), the FRM size would be 51.1\,m and the albedo
0.16; (iii) the ITM reproduces the measured fluxes with a sphere of 53.7\,m diameter (albedo
of 0.14) and a surface temperature of only 207.4\,K. The corresponding model fits to the
six MIRI data points are shown in Fig.~\ref{fig:ratios3}, lower panel).

The more detailed Monte Carlo runs for the NEATM (see Fig.~\ref{fig:montecarlo_neatm}),
the FRM and ITM (both in Fig.~\ref{fig:montecarlo_frm_iso}) confirm our best-fit
values from Fig.~\ref{fig:ratios3}. Based on the Monte-Carlo simulations, the ITM model gave 206.8
$^{+4.6}_{-4.3}$ 4\,K and a diameter of 53.5$^{+3.4}_{-3.3}$\,m (1-$\sigma$ ranges), the FRM produced
50.4 $\pm$ 0.4\,m, and the NEATM led to a best beaming parameter $\eta$ of
3.07$^{+0.31}_{-0.30}$, with a corresponding diameter of 60.5 $\pm$ 3.9\,m.
All three solutions are perfectly compatible with the measured MIRI fluxes.

Using the combined OBS~5 and OBS~6 fluxes (top three lines in Table~\ref{tbl:miri_obs})
produces almost identical size-albedo values, but with lower reduced $\chi^2$ values
of 0.83, 0.97, and 0.48, for the NEATM ($\eta$ = 3.06), the FRM, and the ITM, respectively.
The measured F1500W/F1000W flux ratio from the combined observations is 3.06 $\pm$ 0.13.
This ratio is well reproduced by the NEATM predictions
and perfectly in line with the ITM ratio of 3.02, but outside the predicted FRM ratio of 2.85.

\paragraph{TPM, spherical shape}

\begin{figure}[h!tb]
\centering
\includegraphics[width=\hsize]{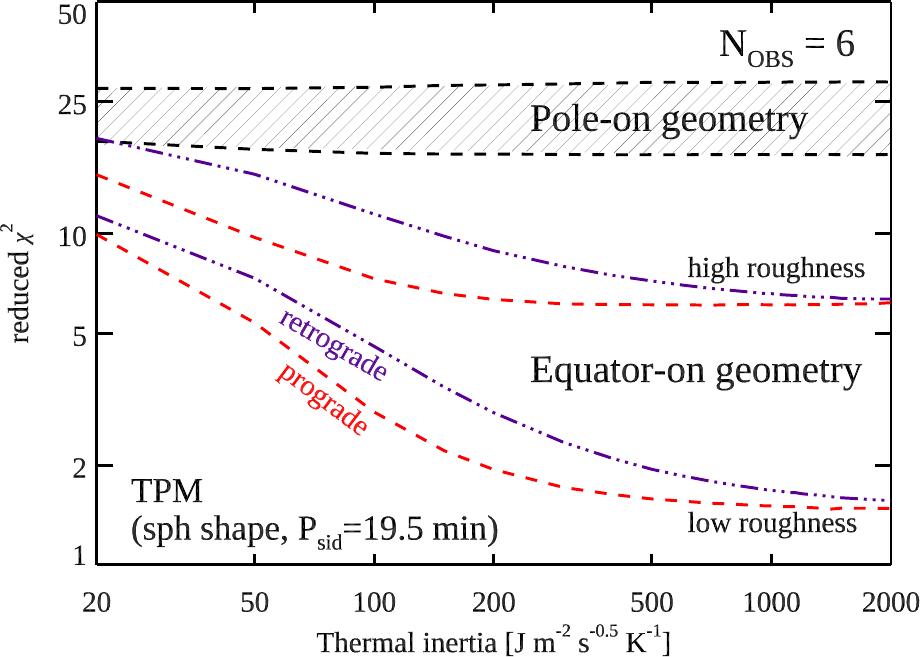}
\includegraphics[width=\hsize]{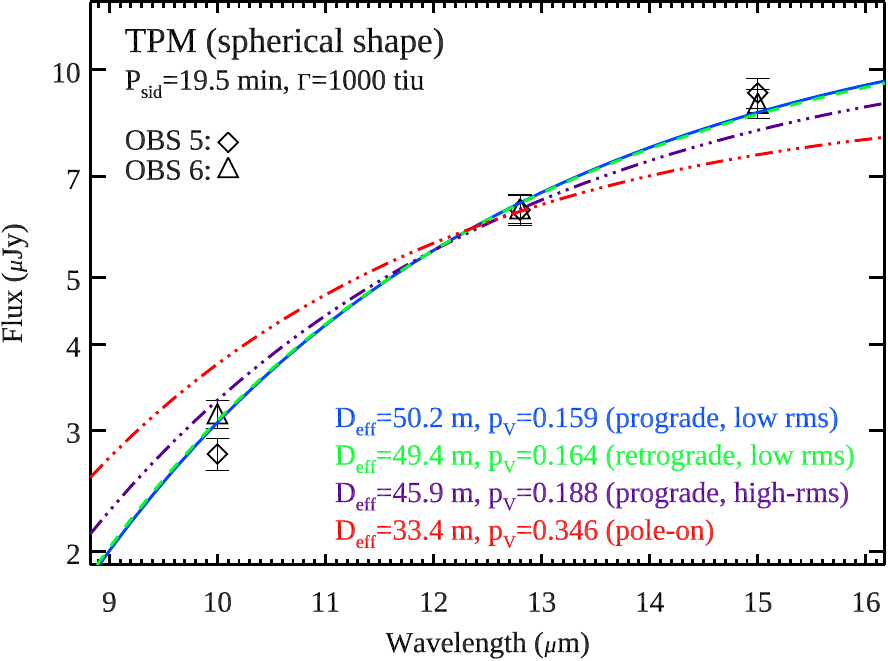}
   \caption{Similar as Fig.~\ref{fig:ratios3}, but now for the TPM (spherical shape model).
            Top: Reduced $\chi^2$-values for TPM calculations using
            a spherical shape solution (with P$_{sid}$=19.5\,min) and a wide range of thermal
            inertias, surface roughness levels, and pole-on and equator-on spin poles.
	    Bottom: OBS~5 and OBS~6 calibrated fluxes (shown as diamonds and triangles)
            together with the best-fit TPM solutions.
      \label{fig:ratios4}}
\end{figure}

For our TPM calculations, we assumed a rotation period of 19.5\,min\footnote{We note that estimates
of the rotation period were already available very shortly after the discovery of the object.}, 
combined with different spin-pole orientations (pole-on and equator-on with either pro- or
retrograde rotation) as seen from JWST on 26 March 2025. In addition to the spin pole, we considered
a wide range of thermal inertia from 0 to 2000\,J\,m$^{-2}$\,K$^{-1}$\,s$^{-1/2}$, going from an
extremely low-conductivity, fine-grained regolith (or extremely porous surface) up to a highly
conductive bare-rock surface. We also varied the surface roughness from very smooth (low mean
surface slopes) to very rough (very high mean surface slopes). The roughness levels are implemented
via hemispherical segmented craters where we modify the width-to-depth ratios and the fraction
of surface covered by these craters \citep[similar to the description in][]{Hanus2015}.
For a given spin-pole and roughness level, we then have only two degrees of freedom in the TPM
fitting to the data: the size and the thermal inertia. The corresponding reduced $\chi^2$ values
are shown in Fig.~\ref{fig:ratios4} (top). Just based on this simple TPM application, it is
possible to rule out a pole-on geometry during the JWST measurements. It became also clear that
a low roughness (or smooth from the thermal point of view), and a high thermal inertia (above about
200\,J\,m$^{-2}$\,K$^{-1}$\,s$^{-1/2}$ for a prograde or above 500\,J\,m$^{-2}$\,K$^{-1}$\,s$^{-1/2}$
for a retrograde rotator) is needed to obtain a good fit to the multi-band MIRI data.
In order to reproduce the high F1500W/F1000W flux
ratio of 3.06 $\pm$ 0.13 the thermal inertia in the equator-on TPM has to be pushed to values of
1000\,J\,m$^{-2}$\,K$^{-1}$\,s$^{-1/2}$ or higher.
The derived radiometric sizes are 50.2\,m (prograde)
and 49.4\,m (retrograde), both calculated at an assumed thermal inertia of 1000\,J\,m$^{-2}$\,K$^{-1}$\,s$^{-1/2}$
and a low surface roughness.

It is important to note that the MIRI fluxes constrain mainly the object's
projected size at the time of the JWST measurements (axis-ratio a/b for
a perfect equator-on viewing geometry), the third dimension (expressed as the
b/c ratio) remains unconstrained in this case.
The geometric albedo is connected to size and absolute magnitude $H$ via:
$log_{10} p_{V} = 6.259 - 2\,log_{10} D_{eff} - 0.4 H$. As YR4's H-mag is 
based on multi-aspect measurements \citep{Devogele2026}, the proper translation
into the object's true geometric albedo also requires an estimate of the effective
3D diameter.

\subsection{TPM, ellipsoidal and convex shapes}
\label{sec:tpm_analysis}

\begin{figure}[h!tb]
\centering
\includegraphics[width=\hsize]{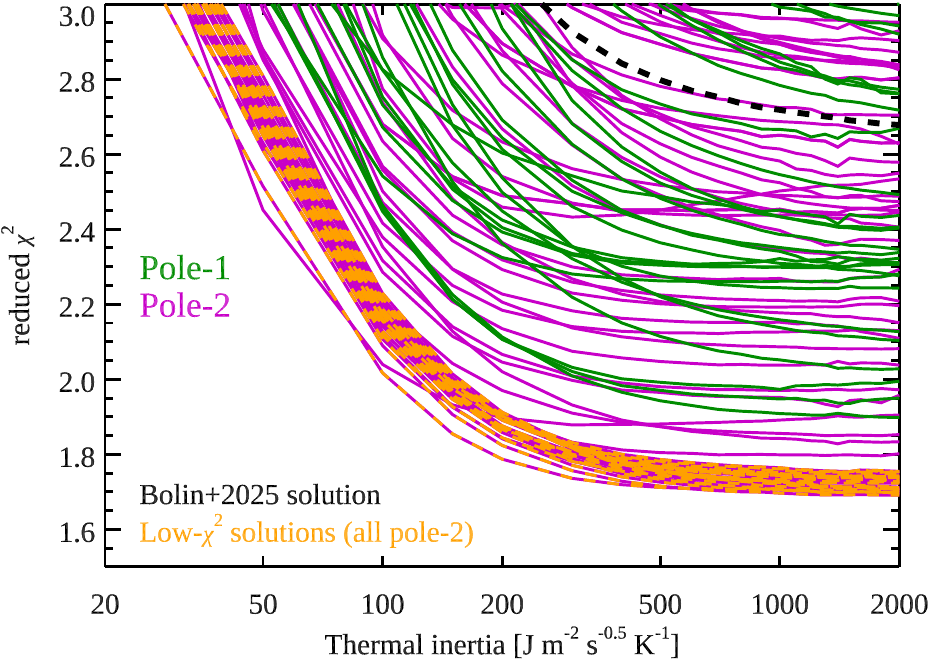}
\includegraphics[width=\hsize]{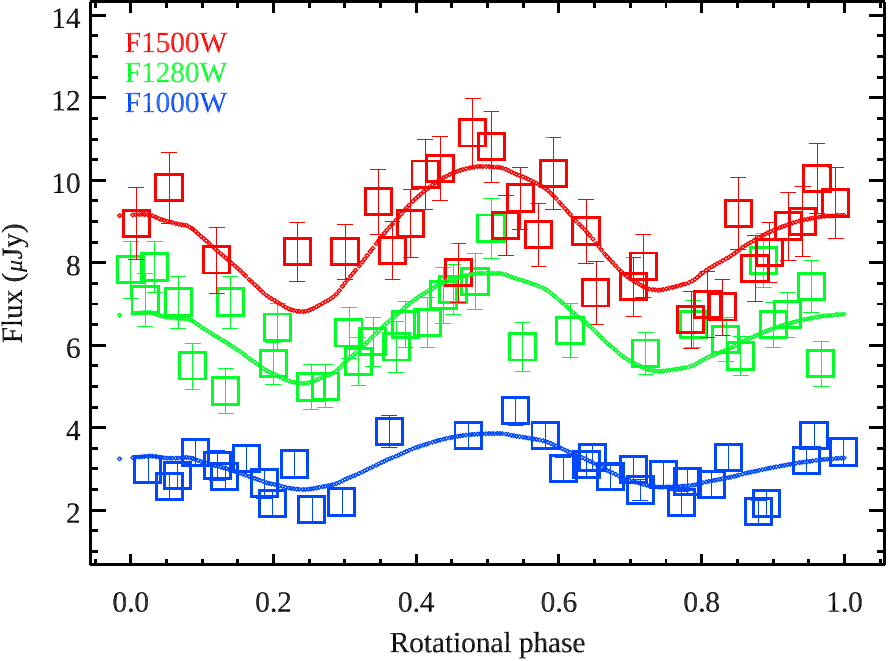}
   \caption{TPM fitting of convex spin-shape solutions to the MIRI data.
            Top: Testing of pole~1 (green lines) and pole~2 (blue lines) solutions.
            Only pole~2 solutions (red lines)
            reach acceptable $\chi^2$ levels. The pole~1 solutions can
            reproduce the absolute flux levels of the MIRI measurements, but
            fail in reproducing the thermal light curves.
            Bottom: Each data point covers about 75 s
            (or 6\% of the rotation period). The three-band thermal light curve is
            clearly visible. Zero rotational phase is identical with the beginning
            of OBS~5 (26 March 2025 05:13:17 UT). The solid lines show the best-fit
            TPM prediction for a minimum-$\chi^2$ pole~2 shape solution, an effective
            size of 60.2\,m (size of an equal-volume sphere), a geometric albedo of 0.11,
            combined with a low surface roughness (2$^{\circ}$ r.m.s.\ of surface slopes),
            and a thermal inertia of 1000\,J\,m$^{-2}$\,K$^{-1}$\,s$^{-1/2}$.
      \label{fig:allfluxes}}
\end{figure}

The LCI in Sect.~\ref{sec:spinshape} produced a range of 
spin-shape solutions, grouped in two spin-pole islands and characterized by
different a/b and b/c axis ratios (see Fig.~\ref{fig:spinshape}).
When implemented in a TPM code, the resulting convex shape models (or approximated ellipsoidal
analog) allow the calculation of thermal light curves. The two thermal light curves
(in amplitude and structure), are highly sensitive to the adopted spin-shape
solution and to the assumed thermal inertia and surface roughness.
However, for testing such predictions, we had to switch from the fluxes 
obtained from the OBS~5 and OBS~6 measurements to fluxes extracted from
the images based on individual MIRI integrations (Fig.~\ref{fig:allfluxes}, bottom panel). Each of these integrations
(see description in Sect.~\ref{sec:miri_obs} and the fluxes in Table~\ref{tbl:miri_obs56_indint})
covers about 75\,s, which corresponds to approximately 6\% of the object's
rotation period. We also tried to use the individual dither images
(with integration times of 308\,s), but each flux covers then about 26\% of 
the object's rotation period, which limits the testing of spin-shape solutions
considerably.

Before applying the calculated spin-shape solutions from Sect.~\ref{sec:spinshape},
we started by simple ellipsoidal shape models with $a$/$b$ between 1.0 and 1.5, with
$b$/$c$ = 1.0. The lowest $\chi^2$-solution ($\chi^2 \sim 1.8$) is found for an $a$/$b$
ratio around 1.3, resulting in an effective size between 52 and 53\,m (and geometric
albedo of $p_V$ = 0.14), and high thermal inertia ($\Gamma > 500$\,J\,m$^{-2}$\,K$^{-1}$\,s$^{-1/2}$).
The axis ratio $b$/$c$ is not well
constrained by the MIRI data as the object was seen nearly equator-on. However,
the $b$/$c$ ratio is very important for the final effective size. Taking typical $b$/$c$ ratios
$\sim$1.4 from the various pole~2 solutions in Sect.~\ref{sec:spinshape} (and $a$/$b$ $\sim$1.3)
push the radiometric effective size (size of an equal-volume sphere) to 59-61\,m ($p_V$ = 0.11).
The two ellipsoidal shape models, with $a$/$b$ = 1.3, $b$/$c$ = 1.4 or with $a$/$b$ = 1.3, $b$/$c$ = 1.0,
combined with the spin pole 2, fit the measured thermal light curve very well, and both indicate
high thermal inertia values above 500\,J\,m$^{-2}$\,K$^{-1}$\,s$^{-1/2}$.

\begin{figure*}[h!tb]
\centering
\includegraphics[width=8cm]{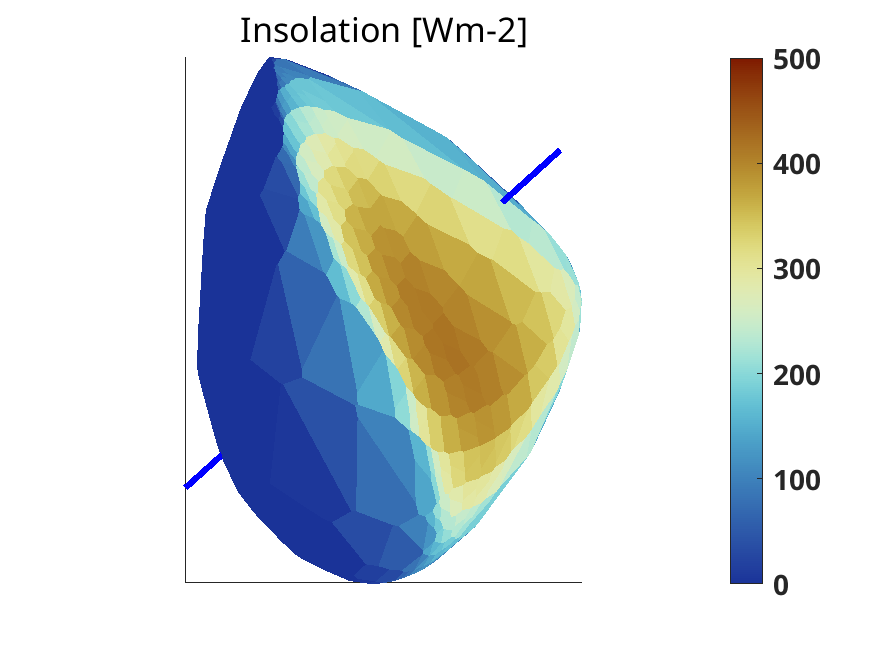}
\includegraphics[width=8cm]{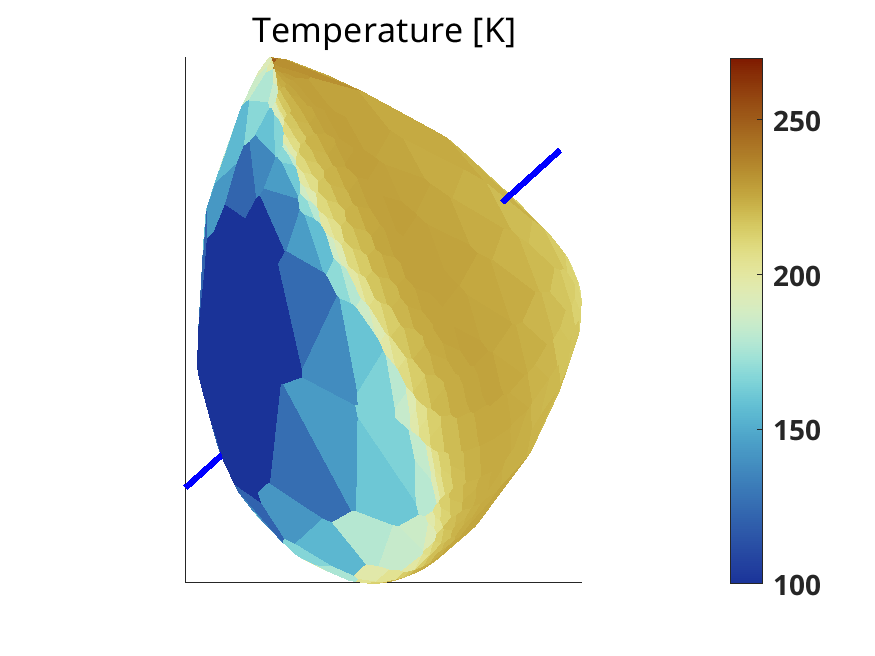}
   \caption{Our best-fit pole 2 solution as seen by JWST on 26 March 2025. Left:
            Solar insolation [W\,m$^{-2}$] at r$_{helio}$ = 1.81\,au and a phase
            angle $\alpha$ = -28.5$^{\circ}$. Right: Surface temperature (in [K])
            assuming a thermal inertia of 1000\,J\,m$^{-2}$\,K$^{-1}$\,s$^{-1/2}$ and a low surface roughness.
            The pole 2 rotation axis is shown in an ecliptic sky-projected 
            viewing geometry.
      \label{fig:insol_temp}}
\end{figure*}

We compare the measured thermal light curve with thermal light-curve predictions
using the available spin-shape solutions. However, based on the available light curves,
which were mostly taken several weeks before the MIRI measurement, we had to leave
the object's zero rotational phase as a free parameter to account for rotational shifts
due to the uncertainties in the light-curve-based rotation period (we note that the 
object rotates more than 500 times per week). For instance, for the spin-shape solution
published by \citet{Bolin2025}, we had to shift the zero rotational phase by
15$^{\circ}$ (corresponding to about 50\,s) to match the phases of the MIRI
light curves.

We analyzed a wide range of possible pole~1 spin-shape solutions, including the
one by \citet{Bolin2025}, but the tested cases could not reproduce the MIRI observations,
either by showing very different levels of light-curve minima or maxima, or by structures
in the model light curve that are not seen in thermal data. The corresponding $\chi^2$
values are far outside an acceptable level (see Fig.~\ref{fig:allfluxes}, top).
The best-fit (in terms of reduced $\chi^2$, by absolute flux levels, and by
matching thermal light-curve structures in Fig.~\ref{fig:allfluxes}) was found for
spin-shape solutions connected to the pole~2 region.
In Fig.~\ref{fig:allfluxes} we show the corresponding TPM prediction as blue, green,
and red solid lines. The band-to-band flux ratios can only be matched by assigning
a high thermal inertia (in combination with low surface roughness), leading to
unexpectedly low surface temperatures around or below $\sim$250\,K (see Fig.~\ref{fig:insol_temp}).
The corresponding
thermal parameter $\Theta$ is above 50 indicating a large heat transport to the night side.
Above the required thermal inertia higher than about 500\,J\,m$^{-2}$\,K$^{-1}$\,s$^{-1/2}$, the object starts to reach
a steady-state temperature distribution along the equator, and the absolute TPM
flux predictions and flux ratios change only very little with increasing thermal
inertia. The derived radiometric size ranges from about 59\,m ($\Gamma$ $\sim$500\,J\,m$^{-2}$\,K$^{-1}$\,s$^{-1/2}$)
to 62\,m ($\Gamma$ $\sim$2000\,J\,m$^{-2}$\,K$^{-1}$\,s$^{-1/2}$) for our pole~2 spin-shape solutions (sizes of an
equal-volume sphere). We repeated the size-albedo calculations for the full range
of 500 spin-shape solutions in the pole~2 region (see Fig.~\ref{fig:spinshape}). 
The high thermal inertia ($>$ 500\,J\,m$^{-2}$\,K$^{-1}$\,s$^{-1/2}$) and low surface roughness (r.m.s.\ of 
surface slopes $<$5$^{\circ}$) is required in all cases.
But not all spin-shape
solutions match the thermal light curves equally well. The 25 best-fit solutions
($\chi^2$ $<$ $\chi^2_{min}$ + n$\sqrt{2\cdot\nu}$, with $\nu$ = 95 (data points)
- 2 (free parameters: thermal inertia \& size) = 93; n = 3, representing 3 standard
deviation in $\chi^2$-space) constrain the spin pole to ($\lambda$, $\beta$) =
(232$^{\circ}$ $\pm$ 1$^{\circ}$, -11$^{\circ}$ $\pm$ 3$^{\circ}$), the
size of an equal-volume sphere to 60.8 $\pm$ 1.8\,m, and the geometric albedo to 0.11$^{+0.04}_{-0.03}$.

\section{Discussion}
\label{sec:dis}

\subsection{Applicability of $\eta$-based thermal models}

The MIRI fluxes of YR4 open the door to very detailed radiometric studies.
A "standard" NEATM analysis for each of the MIRI single band fluxes would lead
to much smaller size solutions. The published beaming parameter for 
several hundred small ($<$10\,km) Mars-crossing asteroid was found to be
$\eta$ = 1.2 $\pm$ 0.2 \citep{Ali-Lagoa2017}. Applying this solution to YR4 would
result in a size range of about 34 -- 40\,m (smaller size at 10.0\,$\mu$m, larger
size at 15.0\,$\mu$m). Even smaller sizes are found when applying the
$\eta = (0.761 \pm 0.009) + (0.00963 \pm 0.00015)\,\alpha$ relation that
was found by \citet{Mainzer2011ApJ743} for 313 NEAs. YR4's phase angle $\alpha$
was -28.5$^{\circ}$ during the MIRI observations leading to a beaming parameter
$\eta$ = 1.04 and the corresponding size range is 32 -- 37\,m. These two size ranges 
would be below the critical threshold of 50\,m for considering active mitigation
planning\footnote{Based on the UN Space Mission Planning Advisory Group (SMPAG):
\url{https://www.unoosa.org/oosa/sk/ourwork/topics/neos/smpag.html}}. For potentially
problematic objects, it is therefore necessary to obtain multi-band MIR data
and to solve for the object's beaming parameter $\eta$ to constrain its size.

The question now is how abundant such high-$\eta$ objects are.
\citet{Mainzer2011ApJ743} found a number of NEAs with high beaming parameters.
They hypothesized that these objects have a high thermal inertia and/or 
rapid rotation. For (1865) Cerberus with a "normal" rotation period of 6.810\,h,
they determined $\eta = 2.94 \pm 0.03$ and speculated about nonzero
temperatures on the night side of the asteroid in contradiction to NEATM assumptions.
\citet{Mainzer2014ApJ784} studied a sample of small NEAs observed by WISE.
Among the 106 NEA, most of them having a sub-kilometer size, they listed seven with
a beaming parameter $\eta \ge 2.5$ and only for one case they saw the need to switch
from the NEATM to the FRM.
Another high-$\eta$ object was presented by \citet{AliLagoa2014}. They found
$\eta$=2.2 $\pm$ 0.4 for the sub-kilometer size NEA (341843) 2008 EV$_{5}$. 
But with a rotation period of 3.725\,h, an equivalent diameter of 400 $\pm$ 50\,m
\citep{Busch2011}, and a thermal inertia of 110$^{+40}_{-12}$\,J\,m$^{-2}$\,K$^{-1}$\,s$^{-1/2}$ \citep{Jiang2019},
2008 EV$_{5}$ does not look to be a very special case in the thermophysical context.
\citet{Ali-Lagoa2017} found only very few objects (maybe around 1\% of the
shown objects in Fig.~3) with $\eta > 2.5$, but they did not follow up on these
objects or the interpretation of the findings.
\citet{Masiero2018} looked at the physical properties of 116 NEAs detected by NEOWISE
during the first three survey years. Six sub-kilometer objects are given with with a 
fitted beaming parameter $\eta \ge 2.5$, but none with a repeated two-band fitted $\eta$ of
larger than 2.8.

One reason for finding high-$\eta$ values can be related to shape effects: if
one band is measured at thermal light-curve minimum (e.g., when seeing the minimum
cross-section) and the second band close to light-curve maximum, then the derived
spectral slope is strongly influenced by the object's light curve.
As a consequence, the $\eta$-determination, which tries to match the band-to-band
flux differences by temperature effects, is no longer reliable.
Also, the combination of MIR data taken at
very different times (different phase angles, different heliocentric distances)
leads to problematic $\eta$ solutions, and to very uncertain
radiometric diameters. However, YR4's very high $\eta$-value is based on 
rotationally-averaged fluxes and flux ratios, all taken at the same observing
geometry. In this context of MIR asteroid surveys and the study of individual
objects, the YR4 case with the extreme $\eta$ appears highly unusual.
The NEATM results are in general not well tested for high-$\eta$ cases.
The handling of phase angles, by ignoring emission from the non-illuminated
surface, leads to a systematic overestimation of the size for increasing
phase angles \citep[see also][and discussions therein]{Delbo2003}.
This size overestimation can be seen for the top-shaped
NEA 2008 EV$_{5}$ where \citet{AliLagoa2014} found a NEATM ($\eta$ = 2.2 $\pm$ 0.4)
size of 470 $\pm$ 70\,m, while the radar and and more detailed TPM studies gave
sizes below 400\,m \citep{Busch2011,AliLagoa2014,MacLennan2022b}.

In our case here, with observations at $\alpha$ = -28.5$^{\circ}$,
the NEATM ($\eta$ = 3.1) application leads to a size of about 60\,m, without
the phase angle correction ($\alpha$ = 0$^{\circ}$), the size would be 56\,m,
at a "half-Moon" phase angle of $\alpha$ = 90.0$^{\circ}$, the resulting size
would already be above 100\,m.

The FRM and the TPM using a spherical shape (or an elliptical shape with $b$/$c$ = 1.0)
produce radiometric diameters of 51 -- 52\,m, while the best-fit NEATM gives 61 $\pm$ 3\,m
(see Table~\ref{tbl:dpv_summary}). The discrepancy is only partly related to the
NEATM handling of phase angles by considering only the surface region, which is directly
illuminated by the Sun. Another aspect is the surface temperature
distribution. In NEATM, the temperatures decrease with the angular distance $\varphi$
from the sub-solar point by $cos^{1/4}\,\varphi$. In the FRM, we
see high temperatures along the entire equatorial band. The temperature decreases
with latitude $\beta$ by $cos^{1/4}\,\beta$ from the equator to the poles.
The different temperature distributions could potentially also cause an overestimation
of the NEATM size in case of high $\eta$ values.
Our YR4 analysis therefore confirms that NEATM applications are
problematic and should be taken with care for
the following cases: (i) for high $\eta$ solutions (e.g., $\eta$ > $\sim$1.5 for low
phase angle measurements); (ii) for objects with known high thermal inertia;
(iii) for fast rotators ($P_{sid} <$ 2\,h), and, (iv) for objects seen under large phase angles
\citep[see also discussions in, e.g.,][]{Harris2002,Delbo2003,Harris2006,Wolters2009,Hanus2015,Mommert2018}.

\subsection{FRM and ITM approximation for high-$\eta$ objects}

\citet{Lebofsky1978} introduced the FRM concept for objects that seem to have
a bare rock, that is, high thermal inertia, possibly combined with a fast rotation
(and expected to show a large beaming parameter; see, e.g., \citealt{Harris1998a,AliLagoa2014}).
A fast rotator with a high thermal inertia would be nearly isothermal along the
equator, with temperatures decreasing towards the poles due to decreasing
insolation. In fact, the FRM size of around 50\,m is significantly lower than
the NEATM size, and very close to the radiometric sizes coming from the more
sophisticated TPM calculations using a spherical or ellipsoidal shape (with $b$/$c$=1.0).

One way to distinguish between the various models would be
by a peak temperature determination. The 3 best-fit solutions have:
(i) NEATM ($\eta$ =3.06): $T_{SS}$ = 223.0\,K; (ii) FRM: $T_{SS}$ = 219.7\,K;
(iii) ITM: $T_{iso}$ = 207.4\,K; (iv) TPM solutions: $T_{max}$ (TPM) $<$250\,K,
(depends on thermal inertia, surface roughness, spin-pole orientation, surface
features). But the wavelength coverage in our MIRI data is not sufficient as
the 10 -- 15\,$\mu$m data are more indicative of the disk-averaged temperatures,
and only a shorter-wavelength measurement, for example, in the F560W or F770W bands,
would have helped to disentangle peak from disk-averaged temperatures.
Another way would be to include measurements in a different epoch and to follow
the object over a wider phase angle range (see also Sect.~\ref{app:predictions}).
The good-quality FRM and ITM fits to the MIRI data indicate that
the NEATM assumption of the surface temperature distribution is not appropriate
for YR4.

The diurnal thermal parameter $\Theta$ (see Appendix~\ref{app:thermal_concepts})
can also help to do a reality check of temperature assumptions in the
various models.
For YR4, using $\sqrt{\omega} = \sqrt{2\,\pi/(19.46341\cdot 60\,s)} \approx 0.073$,
we find thermal parameters $\Theta$ in the range 6 -- 10 (in case of a low-conductivity
lunar-like regolith with $\Gamma\sim$ 50-300\,J\,m$^{-2}$\,K$^{-1}$\,s$^{-1/2}$), and $\Theta$ = 60 -- 160 (for
a rockier, high-conductivity surface with $\Gamma$ $\gtrsim$ 500\,J\,m$^{-2}$\,K$^{-1}$\,s$^{-1/2}$).
Typically, objects with $\Theta$ $\gtrsim$1 are already considered as fast
rotators from the thermal point of view.

The diurnal temperature amplitude scales roughly as 1/$\sqrt{1 + \Theta^2}$
and the surface temperature lags local noon by arctan($\Theta$). Once $\Theta \gtrsim 3$,
the diurnal swings fall roughly as $1/\Theta$, which means that the object spins so
fast that each spot effectively `feels' the time-averaged sunlight.
In this case, the differences between nighttime and daytime temperatures, described
by $\Delta\,T$/$T_{0}$, are very small; we find $\Delta\,T$/$T_{0}$ values below $\sim$4\% for
a low-inertia ($\Gamma <$ 100\,J\,m$^{-2}$\,K$^{-1}$\,s$^{-1/2}$) regolith surface ($6 < \Theta < 10$), and below 1\% for
a higher thermal inertia $> \sim$ 200\,J\,m$^{-2}$\,K$^{-1}$\,s$^{-1/2}$ and/or rockier surface ($\Theta > 25$), meaning
that the night-side temperature is only a few Kelvin lower than the day-side temperature
(less contrast for higher $\Theta$ values). And, the contrast shrinks further away from
the equator, because the diurnal forcing weakens with the cosine of the latitude.
This means, independent of the object's true surface properties, YR4 falls well within 
the FRM regime, and emission also from the non-illuminated part of the
surface has to be taken into account.

The thermal skin depth $\delta = \sqrt{2\,\kappa/\omega} = \sqrt{2\,k/(\rho c \omega)}$
(see Sect.~\ref{app:thermal_concepts} and discussions in \citet{Spencer1989})
is only a few mm for a regolith-covered surface, and a few cm for a rockier surface.
This depth is still tiny for a decameter-size object (the daily fluctuation is confined to this
very thin layer, the interior beneath becomes roughly time-steady, but the surface still
sees a strong radiative boundary condition that enforces a latitudinal gradient).
The lateral heat conduction inside the body is too weak to supply the radiative losses
a warm pole would require: the radiative conductance $h_{rad} = 4\,\epsilon\,\sigma\,T^3$
at $T$ $\sim$ 220\,K is $\approx$2.2 W m$^{-2}$ K$^{-1}$, while the effective lateral
conductance of the body is $k$/$L$ ($k$: conductivity $\sim$ 0.05 -- 2 W m$^{-1}$ K$^{-1}$, with the lower value for fine-grained
regolith, and the higher one for rock, $L$: radius of the object, here about 30\,m), which is
$\ll$ $h_{rad}$ (e.g., $k$/$L$ $\sim$0.002 for regolith and $\sim$ 0.07 for rock). The radiation
thus pins the surface temperature locally to the local insolation, not to the interior.
The pole temperatures can approach equator temperatures only if the object were
very small (size comparable to $\delta$) and/or had very high conductivity (metallic
surface with $k$ $\gg$ 10 W m$^{-2}$ K$^{-1}$), so that $k$/$L$ $\gtrsim$ $h_{rad}$, maybe
in combination with strong self-heating (via deep roughness/concavities) so that poles
"see" warm facets. The metallic and roughness conditions cannot be excluded entirely, but
seem to be very unlikely for typical NEAs, which means that the ITM is not applicable.

\subsection{TPM limitations}

The initial very basic approach to use a spherical shape with the known rotation period of 19.5\,min, pointed already to an object having a high thermal inertia and which was seen nearly equator-on (see Sect.~\ref{sec:radiometric}). But without the known rotation period, the TPM analysis would remain very uncertain.
Adding the spin-pole orientation (which confirmed the initial TPM finding of an equator-on viewing geometry) to the calculations,
allowed us to constrain the object's size (or better: the object's cross-section at the time of the observation). The spherical and the ellipsoidal (with $b$/$c$ = 1.0) spin-shape implementations gave sizes in the range between 51 and 53\,m, very similar to our FRM solutions. In this context, it is interesting to see that a simple ellipsoidal shape (with $a$/$b$ = 1.3 and $b$/$c$ = 1.0) fits also the three-band thermal light curve (Fig.~\ref{fig:allfluxes}) equally well. However, $b$/$c$ = 1.0 shape solutions are not compatible with the visual light curves and the LCI calculations in Sect.~\ref{sec:spinshape} and Fig.~\ref{fig:spinshape}. 
The information about the third axis allows us to go from the cross-section, which is well determined by the MIRI data, to a full 3D body. The effective size (of an equal-volume sphere $D$ = 2$\cdot$($a\cdot$$b\cdot$$c$)$^{1/3}$) increases accordingly (the effective size of a body with $b$/$c$ $>$ 1 has to be larger so that the cross-section remains the same).

The MIRI measurements constrain the thermal inertia to values above about 500\,J\,m$^{-2}$\,K$^{-1}$\,s$^{-1/2}$.
These high values are unexpected as \citet{Petkovic2021}, \citet{Fenucci2021},
\citet{Fenucci2023}, \citet{Novakovic2024a}, and \citet{Marceta2025}
found low thermal inertia values for small, fast-rotating asteroids.
Studies of small samples or on individual NEAs by \citet{Delbo2003}, \citet{Mueller2004a}, or \citet{Harris2007}
suggested that high-thermal-inertia, regolith-free surfaces might be uncommon even among sub-kilometer objects.
In addition, WISE- or NEOWISE-based thermal studies \citep{Mainzer2011ApJ743,Mainzer2014ApJ784,Masiero2018} found
only very few small NEAs with extreme beaming parameters requiring a switch from NEATM to FRM calculations.
These studies point to low thermal inertias for most small NEAs, certainly below 500\,J\,m$^{-2}$\,K$^{-1}$\,s$^{-1/2}$,
and, in contrast to our TPM calculations for YR4 (see Sect.~\ref{sec:tpm_analysis}).

With the knowledge of the object's spin pole, it is also possible to look into
seasonal effects. YR4 was heating up during perihelion passage at 0.85\,au in November 2024.
While the diurnal skip depth is very small (in the millimeter to centimeter range), the
seasonal skin depth is between 2 and 3 meters (taking a thermal inertia of 1000\,J\,m$^{-2}$\,K$^{-1}$\,s$^{-1/2}$).
The estimated radiation time scales at high perihelion temperatures would be one to two
weeks, shifting the phase of maximum temperature from late November to mid
December 2024. With the object moving away from the Sun, the seasonal skin depth remains
the same, but radiation time scales increase due to the decreased overall surface temperature.
As a result, the measured surface temperatures would be above an equilibrium temperature
(for the entire path to aphelion), and typical TPM calculations would overestimate the object's
size. We estimated that a 5 -- 20\% size overestimation might be possible for fast-rotating
objects with high thermal inertia on very eccentric orbits.
However, in the case of YR4, the seasonal heat wave can be neglected. During perihelion phase
the Sun has heated only the North-pole region (sub-solar latitude around +81$^{\circ}$)
when taking our strongly preferred pole~2 solution while during the MIRI measurements we
saw the object close to equator-on (solar aspect angle close to 90$^{\circ}$, see Fig.~\ref{fig:spinshape}, top),
with the sub-solar latitude at -14$^{\circ}$ (see Fig.~\ref{fig:insol_temp}).
The heat transported into the subsurface during perihelion had enough time to radiate away. It is still
possible that North-pole temperatures are underestimated by the TPM calculations, but the
MIR MIRI fluxes are dominated by the warmest equatorial regions on the surface and the
size determination is very likely not affected. However, a small flux contribution from
the seasonal heat wave cannot be excluded entirely. If we assume a remaining 10\% maximum
flux contribution from the north pole region, which was heated during perihelion passage about
four months before our MIRI measurements, then our radiometric size would be overestimated
by about 5\%. Including uncertainties from the seasonal heat wave contribution and in
combination with the wide range of pole~2 spin-shape solutions, we obtain a radiometric
size of 60.8$^{+1.8}_{-3.6}$\,m.

\subsection{Yarkovsky and YORP effects}

The Yarkovsky force \citep{Vokrouhlicky1998,Vokrouhlicky1999} mainly causes
a secular drift of the object's orbital semimajor axis $a$. The force from an-isotropic thermal 
emission acts largely along or opposite to the orbital velocity. And, the force in the tangential 
direction changes the orbital energy, which directly changes $a$. How large could such a 
Yarkovsky drift rate $da/dt$ be? Typical drift rates for sub-kilometer NEAs are tens to 
hundreds of meters of orbital drift per year ($|da/dt| \sim 10^{-4} - 10^{-3}$ au/Myr).
We take a typical reference drift rate of $|da/dt| \sim 10^{-4}$\,au/Myr for a 1 km
asteroid at 1\,au \citep{Vokrouhlicky2000}. The drift rate is size dependent ($da/dt \sim 1/D$),
strongly influenced by the thermal parameter ($da/dt \sim \Theta/(1 + \Theta^2)$), and affected
by the object's distance and the orbit eccentricity. With the diurnal thermal parameter being
very large ($\Theta_{diurnal} >> 10$), the diurnal Yarkovsky drift is expected to be very
small. On the other hand, the seasonal thermal parameter is small ($\Theta_{seasonal} < 0.5$)
and right in the regime of optimal efficiency. In addition, the spin axis is close to the orbital
plane and thus makes the seasonal (inward) drift significant and could be on the order of a few
times 10$^{-4}$ au/Myr, corresponding to a change of the object's semimajor axis up
to 1 km over 10 years time. 
YR4 will have close Earth-Moon encounters in 2028 (0.05\,au) and in 2032 (within 22\,900 $\pm$ 800 km 
of the Moon, \citealt{Rivkin2026}) that will lead to an orbit determination with high precision.
The arc extension can provide an opportunity to measure the (seasonally-dominated)
Yarkovsky effect. However, YR4 is also close to Jupiter's 3:1 MMR \cite{Gladman1997}, which strongly
influences the orbital parameters on short time scales (with an expected dynamic lifetime
of $\sim$10$^{6}$\,Myr).

Is the YORP effect relevant? The YORP effect \citep{Rubincam2000} is a torque acting on small bodies
caused by the an-isotropic reflection and thermal re-emission of solar radiation
from an irregular surface. The shape of YR4 is not symmetric, and the recoil
forces from reflected sunlight and emitted thermal photons do not cancel
completely, producing a net torque. Over long timescales this torque can modify
an asteroid's rotation rate and spin-axis orientation (obliquity). For small
asteroids, especially those below a few hundred meters in diameter, YORP can
significantly accelerate or decelerate the spin and drive the spin pole toward
stable orientations near 0$^{\circ}$ or 180$^{\circ}$ obliquity. But all
currently considered pole directions (pole~1, pole~2, pole-solution by 
\citealt{Bolin2025}) are far away from these end-state orientations. The time-scales
to reach such end-states are $\propto$ $D^{2}\,a^{2}$ and, depending on details
of the irregular shape, can be estimated for YR4 to be between 10$^5$ to 10$^6$ years.
However, close encounters with planets (e.g, YR4's encounter with the Earth-Moon
system in December 2032) or collisions can disturb the YORP end states. Also
thermal cracking and regolith migration can alter the YORP torque on timescales
comparable to the YORP evolution itself \citep{Rozitis2013,Bottke2015,Golubov2022}.

\section{Conclusion}
\label{sec:con}

The radiometric solutions for different model concepts are summarized in Table~\ref{tbl:dpv_summary}.
Based on the model fit to the OBS~5 and OBS~6 fluxes, we found a very high NEATM beaming parameter $\eta$ of around 3.1.
At such high values, and in combination with the short rotation period, a large thermal
parameter ($\Theta > 50$) can be assumed. The high $\Theta$-values lead to an almost isothermal equatorial band.
The NEATM concept to attribute
the measured flux only to the illuminated part of the surface is therefore no longer correct and leads to
an overestimation of the object size. The effect is negligible at small phase angles, but causes a severe
bias at large phase angles. We estimated that the NEATM radiometric size is too large by about 10 -- 15\% (see
entry for the NEATM ellipsoidal shape in Table~\ref{tbl:dpv_summary}).
The FRM, ITM, and TPM solutions (spherical shape) all produce a very similar effective size of around
52 $\pm$ 2\,m. This value can be taken as the object cross-section at the time of the JWST observations.
More complex shapes (ellipsoidal shapes with $b$/$c$ $>$ 1.0, or convex shape solutions) increase the effective
radiometric size to 60.8 $\pm$ 1.8\,m. However, this increase in size with respect to spherical shape
solutions cannot be derived from the single-epoch JWST data. It is purely related to information
from the multi-epoch, multi-aspect angle
LCI exercise, which produces constraints for the object's $b$/$c$ ratio (see Fig.~\ref{fig:spinshape}).
When we also consider a maximum flux contribution of 10\% from the seasonal heat wave (see Sect.~\ref{sec:dis}),
the lower size limit might be smaller by a few percent.

\begin{table*}[h!tb]
  \caption{Summary of the derived YR4 properties for the different model concepts and datasets.
      \label{tbl:dpv_summary}}
  \centering
    \begin{tabular}{llll}
      \hline\hline
      \noalign{\smallskip}
       Model   & $D_{eff}$ [m] & $p_V$ & Thermal characteristics \& remarks \\
      \hline
      \noalign{\smallskip}
      $H_{V}$ = 24.14 $\pm$ 0.25\,mag & 25 -- 128\,m         & 0.03 -- 0.50 & full albedo \& $H$-mag range \\
      $H_{mag}$ =23.9 $\pm$ 0.3       & 30 -- 65\,m          & 0.15 -- 0.40 & \citet{Bolin2025} \\
      NEATM, best-fit $\eta$          & 60.6\,m              & 0.11 & $\eta$ = 3.06, $T_{ss}$ = 222.9\,K, $\chi^2_{r}$ = 1.70 \\
      NEATM, Monte-Carlo simulation   & 60.5$\pm$ 3.9\,m     & 0.11 $\pm$ 0.02 & $\eta$ = 3.07$^{0.31}_{-0.30}$ \\
      NEATM, projected ellipsoid      & 61\,m                & 0.11 & $a$/$b$ = 1.3, $b$/$c$ = 1.0, $\eta$ = 3.1 \\
      NEATM, projected ellipsoid      & 72\,m                & 0.08 & $a$/$b$ = 1.3, $b$/$c$ = 1.4, $\eta$ = 3.1 \\
      FRM, best-fit                   & 51.1\,m              & 0.16 & $\eta$ = $\pi$ (default), $T_{equ}$ = 219.7\,K, $\chi^2_{r}$ = 1.80 \\
      FRM, Monte-Carlo simulation     & 50.4 $\pm$ 0.4\,m    & 0.16 $\pm$ 0.01 &  \\
      ITM, best-fit                   & 53.7\,m              & 0.14 & $T_{iso}$ = 207.4\,K, $\chi^2_{r}$ = 1.63 \\
      ITM, Monte-Carlo simulation     & 53.5$^{+3.4}_{-3.3}$\,m,    & 0.15 $\pm$ 0.01  & $T_{iso}$ = 206.8$^{+4.6}_{-4.3}$\,K \\
      TPM, spin \& spherical shape    & 51 -- 52\,m          & 0.15 & $a$/$b$ = 1.0, $b$/$c$ = 1.0, pole~1 \& 2, $\Gamma$ $>$500\,tiu \\
      TPM, spin \& ellipsoidal shape  & 52 -- 53\,m          & 0.14 & $a$/$b$ = 1.3, $b$/$c$ = 1.0, pole~2, $\Gamma$ $>$500\,tiu \\
                                      & 59 -- 60\,m          & 0.11 & $a$/$b$ = 1.3, $b$/$c$ = 1.4, pole~2, $\Gamma$ $>$500\,tiu \\
      \noalign{\smallskip}
      {\bf TPM, spin \& convex shape} & {\bf 60.8$^{+1.8}_{-3.6}$}    & {\bf 0.11$^{+0.05}_{-0.03}$} & {\bf best-fit pole 2 solutions, $\Gamma$ $>$ 500\,tiu, low rms} \\
      \noalign{\smallskip}
      \hline
    \end{tabular}
    \tablefoot{(1) For all model calculations with a spherical shape, we used N$_OBS$ = 6
                   (OBS~5 and OBS~6 in Table~\ref{tbl:miri_obs}), and for ellipsoidal and
                   convex shapes, we used N$_{OBS}$ = 95 (referring to the fluxes extracted
                   from the individual integration maps; see Table~\ref{tbl:miri_obs56_indint}).\\
               (2) In addition to the derived effective diameter $D$ and the geometric albedo $p_V$ (related
                   to $H$ = 24.14 $\pm$ 0.25\,mag), we also list the temperatures $T_{ss}$ (sub-solar), $T_{equ}$
                   (equatorial temperature), $T_{iso}$ (isothermal equilibrium), the NEATM beaming parameter $\eta$,
                   or thermal inertia $\Gamma$ (tiu $=$ J\,m$^{-2}$\,K$^{-1}$\,s$^{-1/2}$) and surface roughness
                   rms of the surface slopes.}
\end{table*}

Overall, the three-band MIRI measurements provided an excellent dataset for a detailed 
TPM study. We found a radiometric size of 60.8$^{+1.8}_{-3.6}$\,m (size of an equal-volume sphere),
a geometric V-band albedo of $p_V$ = 0.11$^{+0.05}_{-0.03}$ (based on $H_{V}$ = 24.14 $\pm$ 0.25),
a high thermal inertia $\Gamma$ $>$ 500\,J\,m$^{-2}$\,K$^{-1}$\,s$^{-1/2}$, and a low surface
roughness (rms of the surface slopes $<$5$^{\circ}$), causing an almost isothermal equatorial band
with peak temperatures at about or lower than $\sim$250\,K. The estimated size error was based on 25 best-fit
pole 2 spin-shape solutions and thermal inertias higher than 500\,J\,m$^{-2}$\,K$^{-1}$\,s$^{-1/2}$, which matches the MIRI thermal light curves
(see Fig.~\ref{fig:allfluxes}).
The radiometric analysis benefited from the
availability of spin-shape solutions that were derived from multi-aspect light-curve measurements.
The preferred pole 2 solution ($\lambda$, $\beta$)$_{ecl}$ = ($\sim$232$^{\circ}$, $\sim$$-$11$^{\circ}$)
has a rotation period of 19.4633\,min. An equal-inertia ellipsoidal approximation of the pole 2
convex shapes give $a$/$b$ ratios between 1.24 and 1.32. The $b$/$c$ ratio is less well constrained
(see Fig.~\ref{fig:spinshape}) with values between 1.4 and 1.5 (best solution $a$/$b$ = 1.28, $b$/$c$ = 1.45).
The LCI places strong constraints on the spin-pole orientation, the rotation period, and on the 3D shape
of the object. In contrast, the single-epoch MIRI measurement determined the absolute cross
section of the object at a moment when YR4 was seen nearly equator-on.
To consolidate the 3D size, a second MIR measurement (with the object seen under a
different aspect angle) would have been favorable. This second MIR measurement was originally foreseen,
but due to technical reasons, only measurements on 26 March 2025 were executed.

YR4 will again be accessible for JWST in 2028 (see Fig.~\ref{fig:prediction}) with fluxes higher by about
one order of magnitude than in March 2025. A second very short window in late December 2028
to early January 2029 would allow us to detect fluxes in the mJy regime, but this is still too faint for
current ground-based MIR capabilities.
The target might be followed on the way to perihelion from August 
to October 2028 over a wider phase-angle range with $\alpha$ from $-$28.6$^{\circ}$ to $-$66.8$^{\circ}$
and r$_{helio}$ from 1.51\,au to 1.09\,au, and then again after perihelion in late December 2028
to early January 2029, over a phase-angle range from +90.8$^{\circ}$ to +40.7$^{\circ}$ and r$_{helio}$ from  
0.99\,au to 1.09\,au. 
Measurements during these favorable JWST windows would allow us to settle model concepts,
to improve the physical and thermal properties of YR4, and to obtain
multiband full thermal light curves to constrain smaller shape features, and study
diurnal and seasonal effects in more detail.
The changing surface temperatures would inform us about surface material or regolith properties.
Additional thermal measurements would also be needed to constrain the object's seasonal Yarkovsky drift.\\

YR4 does not pose an immediate risk for Earth or the Moon in the foreseeable near-term future
\citep{Rivkin2026}, but it remains an attractive target for planetary defense studies and for improving
procedures for the determination of an object's size and thermal properties.

\begin{acknowledgements}
This work is based on observations made with the
NASA/ESA/CSA James Webb Space Telescope. The data were obtained from
the Mikulski Archive for Space Telescopes at the Space Telescope Science
Institute, which is operated by the Association of Universities for Research in
Astronomy, Inc., under NASA contract NAS 5-03127 for JWST; and from the
European JWST archive (eJWST) (\url{https://jwst.esac.esa.int/archive/})
operated by the ESAC Science Data Centre (ESDC) of the European Space
Agency. These observations are associated with program ID.~\#9239 (PI: A.\ Rivkin).
This research made use of Photutils, an Astropy package for
detection and photometry of astronomical sources \citep{Bradley2025}. P.P. was supported by {\it Praemium Academiae} award (No. AP2401) from the Academy of Sciences of the Czech Republic. Program GN-2025A-DD-103 based on observations obtained at the international Gemini Observatory, a program of NSF NOIRLab, which is managed by the Association of Universities for Research in Astronomy (AURA) under a cooperative agreement with the U.S. National Science Foundation on behalf of the Gemini Observatory partnership: the U.S. National Science Foundation (United States), National Research Council (Canada), Agencia Nacional de Investigaci\'{o}n y Desarrollo (Chile), Ministerio de Ciencia, Tecnolog\'{i}a e Innovaci\'{o}n (Argentina), Minist\'{e}rio da Ci\^{e}ncia, Tecnologia, Inova\c{c}\~{o}es e Comunica\c{c}\~{o}es (Brazil), and Korea Astronomy and Space Science Institute (Republic of Korea).
Part of this work was conducted at the Jet Propulsion Laboratory, California Institute of Technology, under
a contract with NASA (80NM0018D0004), and at the University of Helsinki as part of the grants "Research Council of Finland, 353784, 336546, 359893". 
\end{acknowledgements}

\bibliographystyle{aa}
\bibliography{yr4_references}

\begin{appendix}

\section{Gemini North photometry}
\label{app:obs_gemini}

In Table~\ref{tbl:photometry_mag} we list all available measurements of 2024 YR4 taken with the Gemini-North telescope.
Details of the observations and the data reduction are given in Section~\ref{sec:obs}.

\begin{table}[h!tb]
  \caption{Gemini-North photometric measurements. \label{tbl:photometry_mag}}
  \centering
\begin{tabular}{lcc}
\hline
\hline
\noalign{\smallskip}
JD & Mag & Err \\
\hline
\noalign{\smallskip}
2460740.889028 & 25.033 & 0.276 \\
2460740.893495 & 24.824 & 0.338 \\
2460740.895243 & 25.934$^{a}$ & 0.601 \\
2460740.896991 & 25.132 & 0.199 \\
2460740.898738 & 26.048$^{a}$ & 0.454 \\
2460740.900475 & 25.025 & 0.212 \\
2460740.902222 & 25.387 & 0.291 \\
2460740.903970 & 25.212 & 0.217 \\
2460740.905718 & 25.445 & 0.345 \\
2460740.907465 & 25.783$^{a}$ & 0.409 \\
2460740.909201 & 24.931 & 0.154 \\
2460740.910949 & 25.602 & 0.347 \\
2460740.912697 & 25.507 & 0.298 \\
2460740.914456 & 25.239 & 0.210 \\
2460740.916204 & 24.823 & 0.162 \\
2460740.917951 & 25.003 & 0.169 \\
2460740.919687 & 25.341 & 0.271 \\
2460740.921435 & 25.077 & 0.208 \\
2460740.923183 & 24.806 & 0.152 \\
2460740.924931 & 25.143 & 0.212 \\
2460740.926678 & 25.561 & 0.324 \\
\hline
\end{tabular}
  \tablefoot{
  \tablefoottext{a}{Measurement not used in the shape modeling analysis due to high error ($> 0.4$mag).}
            }
\end{table}

\section{NIRCam photometry}
\label{app:obs_nircam}

NIRCam imaging data (in the two broad bands F150W2 and F322W2) of YR4 were taken on 8 and 26 March 2025,
11 May 2025 (proposal ID 9239), and 18 and 26 February 2026 (proposal ID 9441).
The F322W2 NIRCam measurements from 26 March 2025 as well as all measurements taken in May 2025 \citep{Rivkin2025}
and in February 2026 \citep{Rivkin2026} (proposal ID 9441) were only used for astrometric studies as the object was
too faint to extract photometric light-curve information.
From the JWST-NIRCam imaging data (in the two broad bands F150W2 and F322W2), taken on 8 and 26 March 2025 (each
time about 21\,min per band), we extracted the fluxes and their uncertainties via aperture photometry
\citet{Rigby2023}, including the latest corrections for encircled energy
fractions\footnote{\href{https://jwst-docs.stsci.edu/jwst-near-infrared-camera/nircam-performance/nircam-point-spread-functions}{NIRCam Point Spread Functions}}.
Before applying the photometry procedure to YR4, we first verified our reduction scripts
against two faint calibration stars observed with NIRCam to confirm the star's theoretical model
fluxes.

In Table~\ref{tbl:short_long_fluxes} we list the extracted light-curve-relevant JWST-NIRCam measurements of 2024 YR4
taken in March 2025.

\begin{table*}[h!tb]
   \caption{NIRCam fluxes and uncertainties.\label{tbl:short_long_fluxes}}
   \centering
   \begin{tabular}{rccccrcc}
   \hline
   \hline
UT & \multicolumn{2}{c}{F150W2} & \multicolumn{2}{c}{F322W2} & UT & \multicolumn{2}{c}{F150W2} \\
obs mid time & Flux ($\mu$Jy) & Unc. ($\mu$Jy) & Flux ($\mu$Jy) & Unc. ($\mu$Jy) & obs mid time & Flux ($\mu$Jy) & Unc. ($\mu$Jy) \\
\hline
\noalign{\smallskip}
\multicolumn{5}{l}{8 March 2025\tablefootmark{a}} & \multicolumn{3}{l}{26 March 2025\tablefootmark{b}} \\
      \noalign{\smallskip}
21:55:53 & 0.246 & 0.024 & 0.111 & 0.013 & 09:50:12 & 0.1080 & 0.0209 \\
21:57:51 & 0.266 & 0.029 & 0.122 & 0.013 & 09:51:17 & 0.0845 & 0.0256 \\
21:59:49 & 0.252 & 0.031 & 0.123 & 0.010 & 09:52:21 & 0.0981 & 0.0152 \\
22:04:28 & 0.276 & 0.026 & 0.118 & 0.012 & 09:56:07 & 0.1560 & 0.0146 \\
22:06:26 & 0.372 & 0.040 & 0.131 & 0.014 & 09:57:11 & 0.0425 & 0.0283 \\
22:08:24 & 0.334 & 0.031 & 0.153 & 0.020 & 09:58:16 & 0.1180 & 0.0274 \\
22:13:03 & 0.164 & 0.019 & 0.064 & 0.007 & 10:02:01 & 0.0239 & 0.0118 \\
22:15:02 & 0.237 & 0.023 & 0.111 & 0.013 & 10:03:05 & 0.0839 & 0.0155 \\
22:17:00 & 0.277 & 0.025 & 0.111 & 0.011 & 10:04:10 & 0.0681 & 0.0225 \\
22:24:20 & 0.285 & 0.027 & 0.115 & 0.011 & 10:10:26 & 0.1080 & 0.0314 \\
22:26:18 & 0.354 & 0.034 & 0.181 & 0.020 & 10:11:30 & 0.1420 & 0.0160 \\
22:28:16 & 0.318 & 0.031 & 0.140 & 0.016 & 10:12:34 & 0.1510 & 0.0134 \\ \hline
\end{tabular}
   \tablefoot{
    \tablefoottext{a}{YR4 was at $r$ = 1.651\,au, $\Delta$ = 0.791\,au, $\alpha$ = -25.8$^{\circ}$.}
    \tablefoottext{b}{YR4 was at $r$ = 1.812\,au, $\Delta$ = 1.080\,au, $\alpha$ = -28.5$^{\circ}$.}  
    }
\end{table*}

\section{MIRI photometry}
\label{app:obs_miri}

In addition to the individual observations, we also extracted fluxes for each
dither image (not listed here), as well as for each individual
integration (Table~\ref{tbl:miri_obs56_indint}).

\begin{table*}[h!tb]
  \caption{MIRI imaging mode observations of asteroid 2024~YR$_{4}$: photometry from individual
	  integrations (covering 74.9\,s each). \label{tbl:miri_obs56_indint}}
  \centering
    \begin{tabular}{lllrrrclrrr}
      \hline
      \hline
Filter & Dither Pos.,	  &  EXPMID & SNR    & flx	 & err  & &  EXPMID & SNR    & flx	 & err \\
/Band  & Integration &  (MJD)  &	& ($\mu$Jy) & ($\mu$Jy) & &  (MJD) & & ($\mu$Jy) & ($\mu$Jy) \\
      \noalign{\smallskip}
      \hline
      \noalign{\smallskip}
       &	& \multicolumn{4}{l}{OBS 5} & & \multicolumn{4}{l}{OBS 6} \\
      \noalign{\smallskip}
F1280W & 1, 1st &  60760.21799 &  16 &  7.83 & 0.49 &  &   60760.32459 &  18 &  8.06 & 0.45  \\ 
F1280W & 1, 2nd &  60760.21889 &  15 &  7.05 & 0.47 &  &   60760.32549 &  16 &  7.43 & 0.45  \\ 
F1280W & 1, 3rd &  60760.21979 &  10 &  4.89 & 0.46 &  &   60760.32639 &  15 &  7.11 & 0.48  \\ 
F1280W & 1, 4th &  60760.22069 &  14 &  5.56 & 0.39 &  &   60760.32729 &  12 &  5.50 & 0.45  \\ 
F1280W & 2, 1st &  60760.22361 &  14 &  6.56 & 0.47 &  &   60760.33025 &  13 &  6.30 & 0.49  \\ 
F1280W & 2, 2nd &  60760.22451 &  15 &  7.55 & 0.51 &  &   60760.33115 &  12 &  5.96 & 0.50  \\ 
F1280W & 2, 3rd &  60760.22541 &  12 &  5.96 & 0.47 &  &   60760.33204 &  14 &  7.22 & 0.52  \\ 
F1280W & 2, 4th &  60760.22631 &  12 &  6.36 & 0.53 &  &   60760.33294 &  18 &  8.84 & 0.49  \\ 
F1280W & 3, 1st &  60760.22926 &  16 &  6.14 & 0.38 &  &   60760.33587 &  16 &  5.81 & 0.37  \\ 
F1280W & 3, 2nd &  60760.23016 &  17 &  6.50 & 0.38 &  &   60760.33677 &  16 &  6.51 & 0.41  \\ 
F1280W & 3, 3rd &  60760.23106 &  13 &  5.55 & 0.42 &  &   60760.33767 &  18 &  5.74 & 0.32  \\ 
F1280W & 3, 4th &  60760.23196 &  21 &  7.90 & 0.38 &  &   60760.33856 &  18 &  6.74 & 0.38  \\ 
F1280W & 4, 1st &  60760.23492 &  10 &  4.99 & 0.46 &  &   60760.34152 &  15 &  7.05 & 0.47  \\ 
F1280W & 4, 2nd &  60760.23582 &  12 &  5.61 & 0.47 &  &   60760.34242 &  15 &  6.44 & 0.43  \\ 
F1280W & 4, 3rd &  60760.23671 &  16 &  6.51 & 0.40 &  &   60760.34332 &  12 &  5.01 & 0.42  \\ 
F1280W & 4, 4th &  60760.23761 &  18 &  7.36 & 0.42 &  &   60760.34422 &  13 &  6.09 & 0.47  \\ 
      \noalign{\smallskip}
F1000W & 1, 1st &  60760.24160 &  11 &  2.86 & 0.25 &  &   60760.34826 &  12 &  3.11 & 0.26  \\ 
F1000W & 1, 2nd &  60760.24250 &  10 &  2.62 & 0.25 &  &   60760.34916 &  12 &  2.99 & 0.25  \\ 
F1000W & 1, 3rd &  60760.24340 &   8 &  1.97 & 0.25 &  &   60760.35006 &   9 &  2.18 & 0.24  \\ 
F1000W & 1, 4th &  60760.24430 &  13 &  3.21 & 0.24 &  &   60760.35096 &  13 &  3.28 & 0.25  \\ 
F1000W & 2, 1st &  60760.24722 &  11 &  3.26 & 0.28 &  &   60760.35389 &   9 &  2.58 & 0.27  \\ 
F1000W & 2, 2nd &  60760.24812 &  11 &  3.12 & 0.28 &  &   60760.35479 &  10 &  3.09 & 0.31  \\ 
F1000W & 2, 3rd &  60760.24902 &   7 &  2.20 & 0.30 &  &   60760.35568 &   8 &  2.66 & 0.32  \\ 
F1000W & 2, 4th &  60760.24992 &  12 &  3.91 & 0.31 &  &   60760.35658 &   7 &  2.02 & 0.27  \\ 
F1000W & 3, 1st &  60760.25287 &  16 &  3.82 & 0.24 &  &   60760.35954 &  17 &  3.81 & 0.22  \\ 
F1000W & 3, 2nd &  60760.25377 &  15 &  3.28 & 0.22 &  &   60760.36044 &  19 &  4.41 & 0.24  \\ 
F1000W & 3, 3rd &  60760.25467 &  12 &  2.49 & 0.21 &  &   60760.36134 &  14 &  3.01 & 0.21  \\ 
F1000W & 3, 4th &  60760.25557 &  12 &  2.70 & 0.23 &  &   60760.36224 &  13 &  2.81 & 0.21  \\ 
F1000W & 4, 1st &  60760.25852 &  13 &  3.41 & 0.26 &  &   60760.36519 &   7 &  2.17 & 0.28  \\ 
F1000W & 4, 2nd &  60760.25942 &  10 &  2.84 & 0.27 &  &   60760.36609 &  15 &  3.82 & 0.25  \\ 
F1000W & 4, 3rd &  60760.26032 &  11 &  2.81 & 0.26 &  &   60760.36699 &  11 &  2.98 & 0.26  \\ 
F1000W & 4, 4th &  60760.26122 &   8 &  2.17 & 0.26 &  &   60760.36789 &  12 &  3.40 & 0.27  \\ 
      \noalign{\smallskip}
F1500W & 1, 1st &  60760.26537 &  19 & 10.82 & 0.57 &  &   60760.37197 &  14 &  8.96 & 0.63  \\ 
F1500W & 1, 2nd &  60760.26626 &  16 &  8.68 & 0.55 &  &   60760.37287 &  14 &  7.76 & 0.55  \\ 
F1500W & 1, 3rd &  60760.26716 &  15 &  8.77 & 0.58 &  &   60760.37377 &  18 &  8.91 & 0.50  \\ 
F1500W & 1, 4th &  60760.26806 &  13 &  7.42 & 0.57 &  &   60760.37467 &  16 & 10.17 & 0.63  \\ 
F1500W & 2, 1st &  60760.27099 &  14 &  8.89 & 0.64 &  &   60760.37759 &  10 &  6.99 & 0.68  \\ 
F1500W & 2, 2nd &  60760.27189 &  14 &  9.45 & 0.65 &  &   60760.37849 &  12 &  7.86 & 0.65  \\ 
F1500W & 2, 3rd &  60760.27278 &  15 &  9.82 & 0.63 &  &   60760.37939 &  14 &  9.00 & 0.65  \\ 
F1500W & 2, 4th &  60760.27368 &  12 &  8.07 & 0.64 &  &   60760.38029 &  13 &  8.95 & 0.68  \\ 
F1500W & 3, 1st &  60760.27674 &  17 &  9.48 & 0.55 &  &   60760.38334 &  16 &  8.27 & 0.51  \\ 
F1500W & 3, 2nd &  60760.27763 &  17 & 10.15 & 0.59 &  &   60760.38424 &  19 &  8.27 & 0.44  \\ 
F1500W & 3, 3rd &  60760.27853 &  22 & 11.15 & 0.51 &  &   60760.38514 &  17 &  8.30 & 0.49  \\ 
F1500W & 3, 4th &  60760.27943 &  20 &  9.57 & 0.49 &  &   60760.38604 &  22 & 10.29 & 0.48  \\ 
F1500W & 4, 1st &  60760.28236 &   5 &  3.30$^{\star}$ & 0.58 &  &   60760.38899 &  11 &  7.27 & 0.64  \\ 
F1500W & 4, 2nd &  60760.28326 &  13 &  6.93 & 0.54 &  &   60760.38989 &  13 &  7.92 & 0.59  \\ 
F1500W & 4, 3rd &  60760.28415 &  15 &  8.25 & 0.53 &  &   60760.39079 &  11 &  6.62 & 0.58  \\ 
F1500W & 4, 4th &  60760.28505 &  17 & 10.05 & 0.60 &  &   60760.39169 &  13 &  9.20 & 0.67  \\ 
      \noalign{\smallskip}
      \hline
    \end{tabular}
\tablefoot{Four integrations correspond to a single dither image, and four dither positions
	   are taken for a given observation. The first integration of dither position 4 of
           OBS~5 (in the F1500W band) was not used for the radiometric analysis (marked with $^{\star}$).}.
\end{table*}

\section{Thermal model concepts}
\label{app:thermal_concepts}

We applied a range of different model concepts to determine the object's properties
from observations in the MIR \citep[e.g.,][and references therein]{Delbo2015}.

\paragraph{The Standard Thermal Model (STM) and the near-Earth asteroid thermal model (NEATM).}

Both models assume a spherical shape and instantaneous equilibrium between insolation and thermal emission.
The total absorbed solar radiation is given by: $S_{abs} = \pi (D^2/4)\,S(1-A)$, where $D$ is the diameter,
$S$ the solar flux at the asteroid, and $A$ the bolometric Bond albedo (ratio of total scattered solar energy
in all directions and at all wavelengths to the incident energy). Since the Sun's spectral energy
distribution peaks in the visible wavelength range, we can assume that $A = A_V = q \cdot p_V$,
where $q$ is the phase integral, and $p_V$ is the visible geometric albedo. The phase integral $q$ is
given by $q = 0.29 + 0.684 \cdot G$ (with $G$ being the slope parameter in the $H$-$G$ system, see \citealt{Bowell1989}).
In the $H$,$G_{12}$ system, $q = 0.009082 + 0.4061\ G1 + 0.8092\ G2$ \citep{Muinonen2010}.
The absolute magnitude $H$, geometric albedo $p_V$, and the object's effective diameter $D_{eff}$ are
connected via: $log\,p_V = 6.259 - 2 \cdot log\,D_{eff} - 0.4\,H$.\\
The STM \citep{Lebofsky1986, Lebofsky1989} calculates the object's temperature distribution via
$T(\varphi) = T_{ss} cos^{1/4}\,\varphi$, where $\varphi$ is the angular distance from the sub-solar point,
and $T_{ss}$ is determined from $T_{ss} = [(1-A) S/(\eta \epsilon \sigma))]^{1/4}$ ($\epsilon$ emissivity,
$\sigma$ Stefan-Boltzmann constant, $\eta$ beaming parameter), with the temperature on the
night-side ($\varphi > 90^{\circ}$) assumed to be zero. The beaming parameter $\eta$ accounts for the
observed enhancement of thermal emission at small solar phase angles, mainly due to surface roughness
effects \citep[and references therein]{Harris2002}, it is assumed to be 0.756 in the STM.
In case an asteroid is observed under a nonzero phase angle, the surface integrated flux
(from an integration over all surface elements with the given temperatures $T$) is corrected with an
empirical MIR phase coefficient of 0.01\,mag/deg which is based on observations of main-belt asteroids.
The NEATM \citep{Harris1998} is very similar to the STM but uses the beaming parameter either as
a free parameter (determined from a fit to multi-band measurements), a fixed value for specific object types,
or as a phase-angle dependent parameter. And, instead of taking a fixed MIR phase coefficient, the
NEATM calculates numerically the actual thermal flux from all surface elements visible to the observer
and illuminated by the Sun, that is, as seen under the given phase angle.

\paragraph{The Fast Rotating Model (FRM).}

The FRM \citep{Lebofsky1978,Lebofsky1989}, also called the iso-latitude thermal model (sometimes named ILM),
aims for a better characterization of small, irregularly shaped, fast-rotating objects, which possibly also lack a
dusty insulating regolith and have a high thermal inertia. Here, the temperature $T$ only changes with
latitude $\beta$, and half of the thermal emission originates from the night side: $T(\beta) = T_{SS} \cdot cos^{1/4}\,\beta$
and $T_{ss} = [(1-A) \cdot S / (\pi \epsilon \sigma)]^{1/4}$ (corresponding in principle to a beaming parameter $\eta = \pi$,
see also \citealt{Mainzer2011ApJ743} or \citealt{Mainzer2014ApJ784}).

\paragraph{Isothermal Model (ITM).}

The ITM is a very simple model assuming a full heat redistribution on the surface as
described by a single temperature. The surface's equilibrium isothermal temperature can be calculated via
$T_{iso} = [(1-A) \cdot S / (4 \epsilon \sigma)]^{1/4}$. The factor of 4 comes from the ratio
between absorbing area ($\pi\,R^{2}$) and radiating area ($4\,\pi\,R^{2}$), with $R$ being the radius
of the object \citep[see, e.g.,][]{dePater2015}. The bolometric emissivity $\epsilon = 0.9$ is the default
value and applicable to most (silicate mineral) regolith \citep[e.g.,][]{Brown1982}. Alternatively, the isothermal temperature can be left as a free (fitted) parameter: $T_{iso}$.
In both cases, the ITM temperature is homogeneous over the surface (no dependence on the angular-distance from the sub-solar point), and independent of the phase angle under which the object is seen.

\paragraph{Thermophysical Model (TPM).}

The different TPMs \citep[and references therein]{Delbo2015} typically allow to take the object's spin
and shape into account. They consider the true illumination and observing geometry, 1-D heat
conduction into the surface, self-heating and shadowing effects, wavelength- and directional-dependent
emissivity characteristics, and model surface roughness effects. TPM concepts are often used in the
context of space mission targets \citep[e.g.,][]{Mueller2014}. The TPM predictions are more realistic,
but also require a detailed knowledge of an object's physical and thermal properties. Testing and
validation of the TPM concepts have been conducted for objects down to the sub-kilometer size range
(e.g., for the OSIRIS-Rex and Hayabusa2 mission targets, Bennu and Ryugu, respectively). However,
the TPM concepts have not been fully verified for the decameter-size regime, for viewing geometries
under extreme phase angles, or for very fast rotating objects.

\paragraph{Diurnal and seasonal thermal effects.}

The thermal skin depth $l$ is the depth at which the temperature variation has dropped to 1/$e$ ($\sim$37\%)
of its surface amplitude. It is calculated via $l = \sqrt{\frac{\kappa}{\rho c \omega}} = \frac{\Gamma}{\rho c \sqrt{\omega}}$
where $\Gamma$ is the thermal inertia (J\,m$^{-2}$\,K$^{-1}$\,s$^{-1/2}$)\footnote{thermal inertia unit (tiu) in Table~\ref{tbl:dpv_summary}}, $\rho$ the object's
density, $c$ the heat capacity, and $\omega$ the angular frequency of the asteroid.
For the diurnal skin depth, $\omega$ is calculated via ($2\,\pi/P$), with $P$ being the object's rotation period,
while for the seasonal skin depth, $P$ is replaced by the $P_{orb}$, the orbital period.
If we take the assumption of a stony object with moderate roughness we take $\rho$ = 2500 kg m$^{-3}$ and $c$ = 700 J kg$^{-1}$ K$^{-1}$
(for a 200-300\,K temperature regime), we find for YR4 (rotation period 19.5\,min, orbital period 3.99\,yr) a
diurnal thermal skin depth in the millimeter to centimeter regime (higher values for higher thermal inertias),
and a seasonal thermal skin depth of about 0.25\,m to almost 4\,m (again, higher values for higher thermal inertias).

The radiation timescale $t_{R}$ can also be estimated
\citep{Spencer1989}. In addition to the influence of the thermal inertia
and the angular frequency, it is the absolute temperature which controls the cooling:
$t_R = \frac{\Gamma}{\sqrt{\omega} \epsilon \sigma T^3}$. The diurnal time scale at
perihelion (at 0.85\,au on 23 November 2024) would be between about 5\,min up to 1.5\,hours
(higher values for higher thermal inertias). At 1.8\,au from the Sun (during our MIRI
measurements), the diurnal time scales are already between $\sim$15\,min and several
hours, depending on the thermal inertia and the effective surface temperature.

The seasonal time scales are much longer. At perihelion,
we estimated time scales between about one day up to more than two weeks (for very high
thermal inertias). At the 1.8\,au heliocentric distance, the timescales increase
to about 4\,days up to several months, depending again on thermal inertia and
effective surface temperatures. YR4's current aphelion location is at 4.2\,au where
surface temperatures would drop well below 200\,K. The radiation timescales are
then even longer, ranging from two weeks up to more than half a year.

The dimensionless thermal parameter, $\Theta$, characterizes the thermal behavior of an asteroid's surface
under solar radiation \citep{Spencer1989}. It is defined as: $\Theta = \Gamma \sqrt{\omega}/(\epsilon \sigma T^3)$
with $\epsilon$ the emissivity of the surface, $\sigma$ the Stefan-Boltzmann constant (5.67 $\cdot$ 10$^{-8}$ W\,m$^{-2}$K$^{-4}$),
and $T$ the sub-solar temperature \citep{Spencer1990}. Here again, the diurnal thermal parameter requires the object's rotation
period, the seasonal thermal parameter is related to the orbital period.

The Moon temperature is very close to being in equilibrium with Sun-light
and has $\Theta_{diurnal}$ = 0.025, a fast-rotating, high-inertia object might be close to the isothermal-latitude
model (with $\Theta_{diurnal}$ $\rightarrow$ $\infty$), most asteroids are probably somewhere in between and
require thermophysical model considerations when predicting their MIR brightness. As small asteroids tend
to rotate faster \citep{Pravec2008}, a high thermal parameter seems to be very likely for decameter-scale
objects. The calculated diurnal thermal parameters for YR4 range between $\Theta_{diurnal}$ $\sim$5.5 (for a very low thermal
inertia of 100\,J\,m$^{-2}$\,K$^{-1}$\,s$^{-1/2}$) up to $\Theta_{diurnal} > 50$ (for thermal inertias larger
than 500\,J\,m$^{-2}$\,K$^{-1}$\,s$^{-1/2}$). On the other side, the seasonal
thermal parameter is much smaller and ranges from values below 0.01 for a low thermal inertia and
during the perihelion passage, up to about 0.9 for a high thermal inertia during aphelion.
\citet{Petkovic2021}, \citet{Fenucci2021}, \citet{Fenucci2023}, and \citet{Marceta2025} found low thermal inertia
values for small, fast-rotating asteroids which would lower the thermal parameter again.
In addition, a WISE-based thermal study \citep{Mainzer2014} found beaming parameters between 1.0 and
1.5 (far from the FRM beaming parameter of $\pi$) for about 50 near-Earth objects in the size
range between 8 and about 100\,m. Both studies would point to low thermal inertias, certainly well below
500\,J\,m$^{-2}$\,K$^{-1}$\,s$^{-1/2}$, which would mean $\Theta$-values $<$ $\sim$25 for YR4.

\section{Monte-Carlo NEATM, FRM, and ITM size solutions}
\label{app:simple_sizes}

In parallel to the best-fit solutions presented in Sect.~\ref{sec:spherical_results}
we performed Monte Carlo runs \citep[e.g.,][]{Press2007} to estimate the uncertainties of the derived model
parameters. For each realization, the observed fluxes (OBS~5 and OBS~6 fluxes listed in Table~\ref{tbl:miri_obs})
were perturbed by adding random Gaussian deviates with standard deviations equal
to their quoted 1-$\sigma$ uncertainties (as before, we added another 3\% to the flux errors
to take limitations in the background elimination into account). 
The perturbed dataset was then fitted in the same way as the original observations.
Repeating this procedure 10,000 times yields distributions of the best-fit parameters,
from which the median (or mean) values and the 68\% confidence intervals (1-$\sigma$)
were derived. It is important to mention that the Monte Carlo analysis propagates the
measurement uncertainties only. It does not account for systematic uncertainties
arising from the model concept itself.

\begin{figure}[h!tb]
\centering
\includegraphics[width=\hsize]{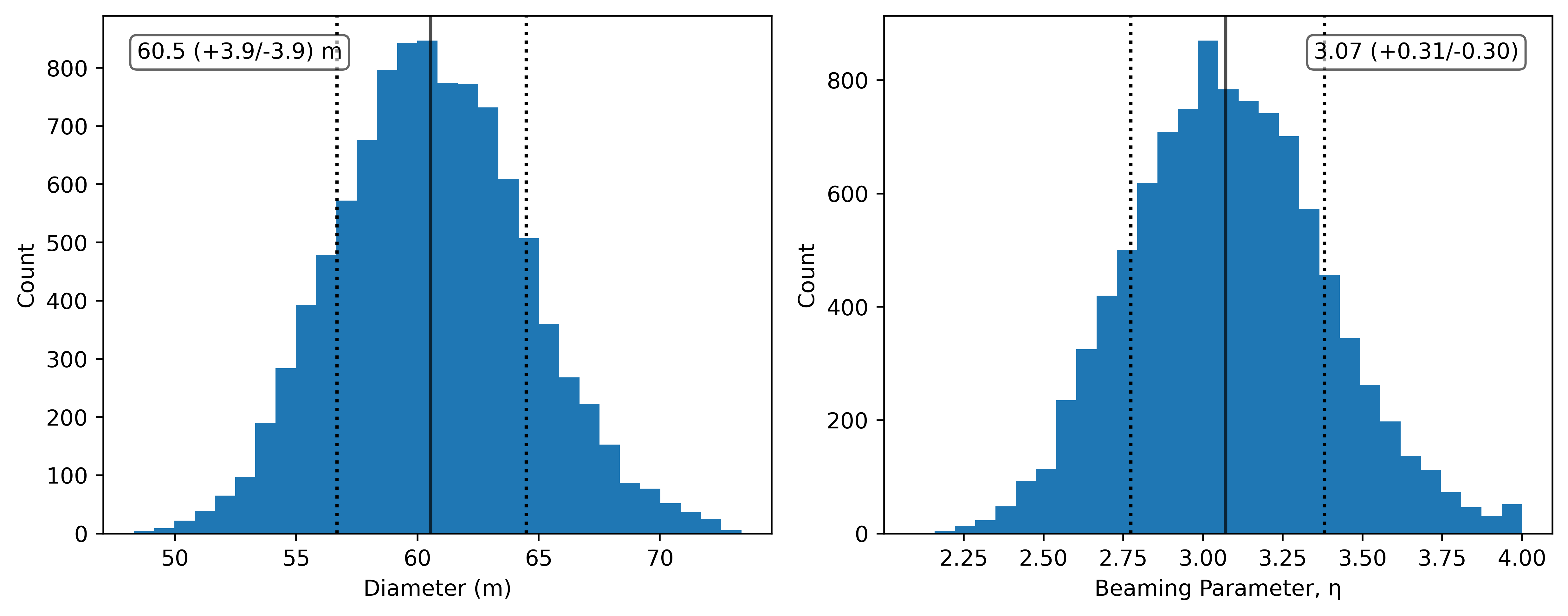}
   \caption{The results of 10\,000 Monte Carlo runs to find the best size-albedo
            solution via the NEATM (with two free parameters: size and beaming
            parameter $\eta$). Left: The diameter distribution. Right: the 
            distribution for the beaming parameter $\eta$.
      \label{fig:montecarlo_neatm}}
\end{figure}

\begin{figure}[h!tb]
\centering
\includegraphics[width=0.48\hsize]{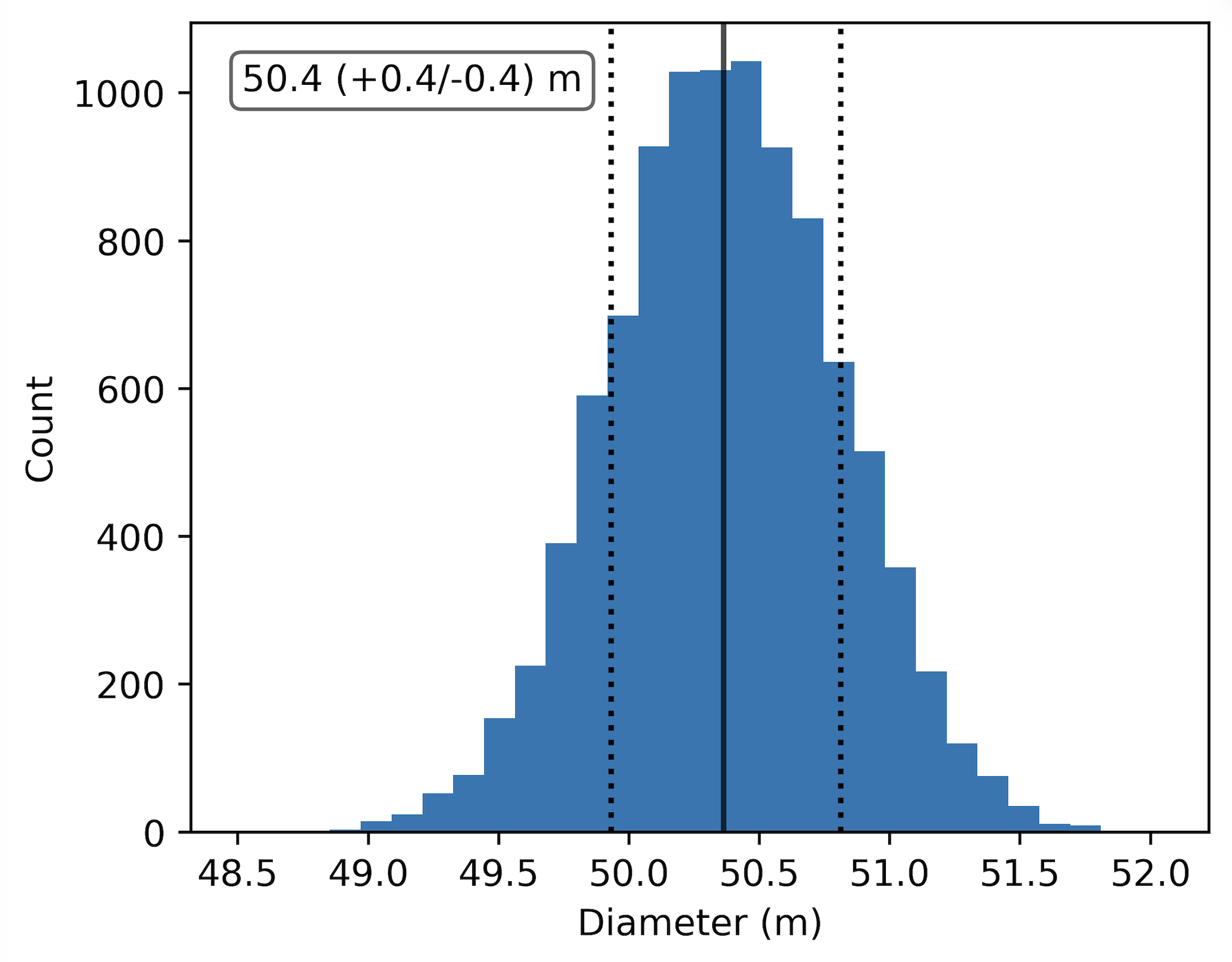}
\includegraphics[width=0.48\hsize]{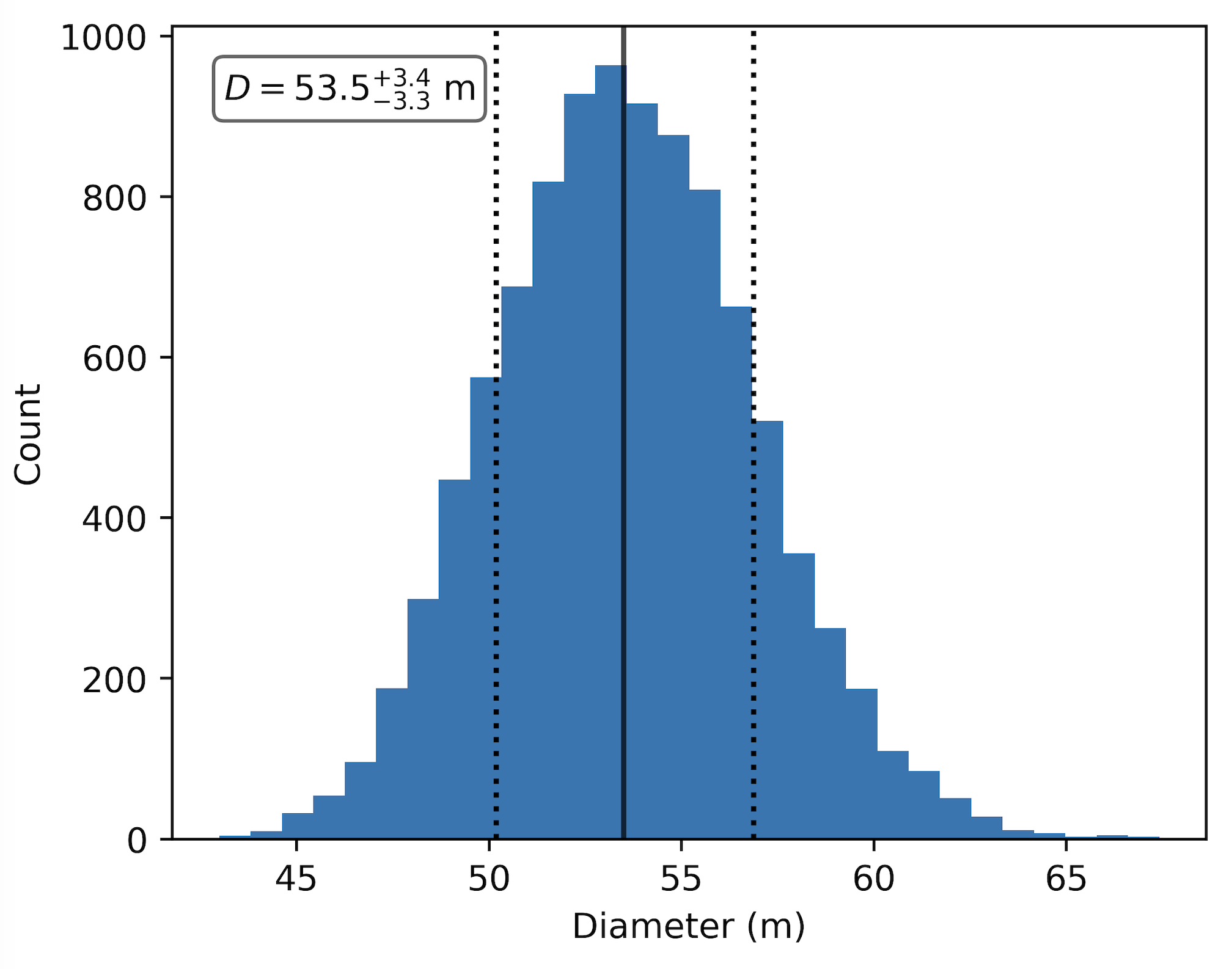}
   \caption{The results of 10\,000 Monte Carlo runs to find the best
            solution in terms of the object's size. Left: for the FRM with only one
            free parameter (size). Right: The ITM solution. 
      \label{fig:montecarlo_frm_iso}}
\end{figure}

The results are shown in Figs.~\ref{fig:montecarlo_neatm} and \ref{fig:montecarlo_frm_iso}.

\subsection{Predictions}
\label{app:predictions}

\begin{figure}[h!tb]
\centering
\includegraphics[width=\hsize]{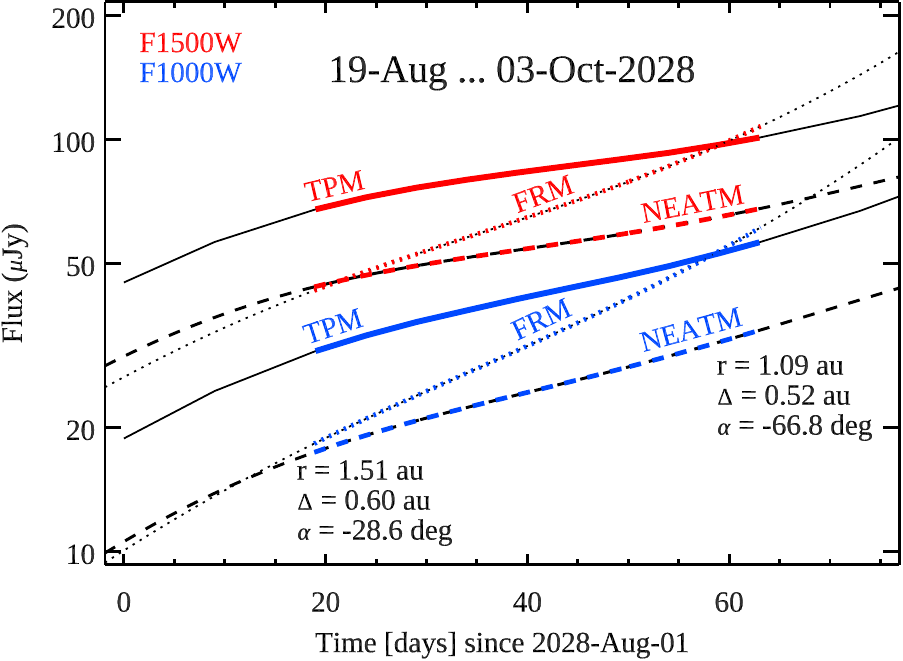}
   \caption{Thermal model predictions for August, September, and October 2028. The colored lines
            (solid: TPM, dashed: NEATM, dotted: FRM) indicate the time period
            when YR4 is inside the JWST accessibility zone. All model predictions
            are based on the best-fit solutions for the MIRI March 2025 data.
      \label{fig:prediction}}
\end{figure}

YR4 reaches aphelion ($Q$ = 4.18\,au) in November 2026, followed by perihelion ($q$ = 0.85\,au) in
November 2028 and a close Earth encounter (about 20 lunar distances) in mid-December 2028. In the
first JWST visibility window in 2028 (10 April - 11 June), the predicted fluxes of 1-5\,$\mu$Jy (10-15\,$\mu$m)
are too low for detailed thermal studies. However, during the 2028 August 19 to October 3 window, its visual
magnitude of around 25 is nearly constant (the decreasing distance is more or less compensated by 
an increasing phase angle), while the MIR flux increases by a factor of 3 (see Fig.~\ref{fig:prediction})
enabling high-SNR MIRI observations. Model predictions (FRM, NEATM, TPM) show measurable 
differences in the F1000W and F1500W bands, despite all fitting the MIRI data presented here from 
March 2025. Multi-band observations in September 2028 would therefore strongly constrain thermal properties.

After the December 2028 encounter, YR4 re-enters JWST visibility (2028 December 23 to 2029 January 4), 
reaching mJy-level MIR fluxes at large phase angles (around 90$^{\circ}$). These fluxes are still
at the limit of current ground-based detection capabilities, but MIRI observations could
probe subsurface properties, separate diurnal and seasonal effects, and provide constraints
on Yarkovsky accelerations in a high-eccentricity orbit ($e$ = 0.66) \citep[see, e.g.,][]{Paoli2026}.

We also investigated detection possibilities for upcoming or planned MIR survey missions
such as NEO Surveyor \citep{Mainzer2023, Masiero2024a, Masiero2024b} and
NEOMIR \citep{Conversi2024, Conversi2026}. Both $\sim$0.5\,m telescopes will be
placed at the Sun-Earth-Moon Lagrangian point L1, about 1.5\,million km away
from Earth in the direction of the Sun, and they plan to operate at MIR wavelengths.
As seen from L1, the calculated 8~$\mu$m flux exceeds the theoretical detection
threshold of 150\,$\mu$Jy about 40 days before the closest Earth encounter, in 2028,
and also in 2032 (while the target, as seen from Earth is still fainter than
magnitude 25). However, NEOMIR's sky accessibility window (solar elongation
between 30 and 70$^{\circ}$) is much better suited to follow the target for most of
the trajectory while in the case of NEO Surveyor (smallest foreseen elongation is 45$^{\circ}$)
the target disappears about 25 and 15 days before the Earth encounter in 2028 and 2032,
respectively \citep{Mueller2026}.
Overall, such fast-rotating, almost isothermal objects,
even if they are only a few 10s of meters in size, are much easier to detect at MIR
wavelengths. The typical brightness decrease with phase angle limits visual observations
as small solar elongations while at thermal wavelengths this is much less of a problem
\citep{Mueller2025}.

\end{appendix}

\end{document}